\documentclass[12pt]{article}

\usepackage{amsfonts}
\usepackage{amsmath}
\usepackage{cite}
\usepackage{epsfig}
\usepackage{latexsym}
\usepackage{paralist}
\usepackage{graphicx}
\numberwithin{equation}{section}
\def\spa#1{\phantom{\fbox{\rule[-#1cm]{0cm}{0cm}}}}

\def\be{\begin{equation}}
\def\ee{\end{equation}}
\def\bea{\begin{eqnarray}}
\def\eea{\end{eqnarray}}

\def\half{{1\over 2}}

\def\nn{\nonumber}

\renewcommand{\thefootnote}{\fnsymbol{footnote}}

\def\bbC{{\mathbb{C}}}

\def\bbZ{{\mathbb{Z}}}
\def\Vh{{\widehat{V}}}
\def\Zh{{\widehat{Z}}}
\def\Nt{{\widetilde{N}}}
\def\cN{{\cal N}}
\def\cO{{\cal O}}
\def\sign{\mathop{\mathrm{sign}}\nolimits}
\def\tr{\mathop{\mathrm{tr}}\nolimits}

\begin{document}

\hfuzz=100pt
\title{{\Large \bf{The Partition Function of ABJ Theory}}}
\date{}
\author{Hidetoshi Awata$^a$\footnote{awata@math.nagoya-u.ac.jp}, 
Shinji Hirano$^b$\footnote{hirano@eken.phys.nagoya-u.ac.jp}, and Masaki Shigemori$^c$\footnote{shige@kmi.nagoya-u.ac.jp}
  \spa{0.5} \\
$^{a}${{\it Department of Mathematics}}
\\ {{\it Nagoya University}}
\\ {{\it Nagoya 464-8602, Japan}}
  \spa{0.5} \\
$^b${{\it Department of Physics}}
\\ {{\it Nagoya University}}
\\ {{\it Nagoya 464-8602, Japan}}
\spa{0.5}  \\
$^c${{\it Kobayashi-Maskawa Institute}}
\\ {{\it for the Origion of Particles and the Universe}}
\\ {{\it Nagoya University}}
\\ {{\it Nagoya 464-8602, Japan}}
}
\date{}

\maketitle
\centerline{}

\begin{abstract}
We study the partition function of the ${\cal N}=6$ supersymmetric
$U(N_1)_k\times U(N_2)_{-k}$ Chern-Simons-matter (CSM) theory, also
known as the ABJ theory. For this purpose, we first compute the partition
function of the $U(N_1)\times U(N_2)$ lens space matrix model exactly.  The
result can be expressed as a product of $q$-deformed Barnes $G$-function
and a generalization of multiple $q$-hypergeometric function.  The ABJ
partition function is then obtained from the lens space partition
function by analytically continuing $N_2$ to $-N_2$. The answer is given
by ${\rm min}(N_1,N_2)$-dimensional integrals and generalizes the
``mirror description'' of the partition function of the ABJM theory, {\it
i.e.}\ the ${\cal N}=6$ supersymmetric $U(N)_k\times U(N)_{-k}$ CSM theory.
Our expression correctly reproduces perturbative expansions and vanishes
for $|N_1-N_2|>k$ in line with the conjectured supersymmetry breaking,
and the Seiberg duality is explicitly checked for a class of nontrivial
examples.
\end{abstract}

\renewcommand{\thefootnote}{\arabic{footnote}}
\setcounter{footnote}{0}

\newpage

%%%%%%%%%%%%%%%%%%%%%%%%%%%%%%%%%%%%%%%%%%%%%%%%%%%%%%%%%%

\section{Introduction}

There has recently been remarkable progress in applications of the
localization technique \cite{Witten:1988ze} to supersymmetric gauge
theories, notably in dimensions $D\ge 3$: In $D=4$ the Seiberg-Witten
prepotential of ${\cal N}=2$ supersymmetric QCD \cite{Nekrasov:2002qd}
was directly evaluated, and the partition functions and BPS Wilson loops
of the ${\cal N}=2$ (and $2^{*}$) and ${\cal N}=4$ supersymmetric
Yang-Mills theories (SYM) were reduced to eigenvalue integrals of the
matrix model type \cite{Pestun:2007rz}, providing, in particular, a
proof of the earlier results on a Wilson loop in the ${\cal N}=4$ SYM
\cite{Erickson:2000af, Drukker:2000rr}. In $D=3$ similar results were
obtained for the partition functions and BPS Wilson loops of ${\cal
N}=2$ supersymmetric Chern-Simons-matter (CSM) theories
\cite{Kapustin:2009kz, Hama:2011ea}, including the ${\cal N}=6$
superconformal theories constructed by
Aharony, Bergman, Jafferis and Maldacena (ABJM)
\cite{Aharony:2008ug}\cite{Aharony:2008gk}. More recently, the
localization technique was further applied to the partition functions of
$5$-dimensional SYM with or without matter \cite{Hosomichi:2012ek,
Kallen:2012va, Kim:2012av}.

The localization method, resulting in the eigenvalue integrals of the matrix model type, allows us to obtain various exact results at strong coupling of supersymmetric gauge theories. In particular, these results provide useful data for the tests of the AdS/CFT correspondence  \cite{Maldacena:1997re} in the case of superconformal gauge theories. For instance,  
the precise agreement of the $N^{3/2}$ scaling between the free energy of the ABJ(M) theory \cite{Drukker:2010nc, Marino:2011nm, triSE} and its AdS$_4$ dual \cite{Aharony:2008ug}\cite{Aharony:2008gk} is an important landmark that shows the power of the localization method in the context of AdS/CFT\@.
Rather remarkably, exact agreements were also found in \cite{Jafferis:2012iv} between the $N^{5/2}$ scaling of 5d superconformal theories and that of their AdS$_6$ duals \cite{Brandhuber:1999np}. Furthermore, the tantalizing $N^3$ scaling of maximally supersymmetric 5d SYM was found in  \cite{Kim:2012av, Kallen:2012zn} in line with the conjecture on $(2,0)$ 6d superconformal theory compactified on $S^1$  \cite{Douglas:2010iu}, despite thus far a lack of the precise agreement with its AdS$_7$ dual.
It should, however, be noted that the utility of the localization method, unlike the integrability \cite{Beisert:2010jr}, is limited to a class of supersymmetric observables, such as the partition function and BPS Wilson loops. On the other hand, the localization method has an advantage over the integrability in that it can provide exact results at strong coupling beyond the large $N$ limit, where the integrability has not been as powerful.

In this paper we focus on the partition function of the ABJ theory, {\it i.e.}, the ${\cal N}=6$ supersymmetric
$U(N_1)_k\times U(N_2)_{-k}$ CSM theory, which generalizes the equal rank $N_1=N_2$ case of the ABJM theory \cite{Aharony:2008gk}.
Over the past few years there has been considerable progress in the study of the partition function and Wilson loops of the ABJM theory, whereas the ABJ case has not been as much understood. The ABJ generalization, for instance, has an important new feature, the Seiberg duality, which, however, lacks a full understanding.  
Besides being a generalization, it has recently been conjectured that
the ABJ theory at large $N_2$ and $k$ with $N_2/k$ and $N_1$ fixed
finite is dual to the ${\cal N}=6$ parity-violating Vasiliev higher spin
theory on AdS$_4$ with $U(N_1)$ gauge symmetry \cite{Chang:2012kt}. Thus
a better understanding of the ABJ theory may provide valuable insights
into the relation between higher spin particles and strings. It is
therefore worth studying the partition function of the ABJ theory in
great detail.
 
As mentioned above, the partition function of the ABJM theory has been well studied. In the large $N$ limit, the planar free energy has been computed, revealing the aforementioned $N^{3/2}$ scaling \cite{Drukker:2010nc, Marino:2011nm, triSE}. In fact, the result in \cite{Drukker:2010nc, Marino:2011nm} is exact in 't Hooft coupling $\lambda=N/k$ and, in particular, confirms a gravity prediction of the AdS radius shift in \cite{Bergman:2009zh}. The planar result is not limited to the ABJM case; Drukker-Mari\~no-Putrov's results include the partition function and Wilson loops of the ABJ theory, and the ABJ version of the radius shift  \cite{Aharony:2009fc} is also confirmed.
In the meantime, beyond the large $N$ limit, the $1/N$ corrections of the ABJM partition function were summed up to all orders by solving the holomorphic anomaly equations of \cite{Bershadsky:1993cx, Drukker:2010nc, Drukker:2011zy} at large $\lambda$ in the type IIA regime $k\gg 1$, and the result turned out to be simply an Airy function \cite{Fuji:2011km}.\footnote{There remains an unresolved mismatch in the $1/N^2$ correction to the AdS radius shift between the field theory \cite{Drukker:2011zy, Fuji:2011km} and the gravity dual \cite{Bergman:2009zh}. On the other hand, quite recently, a one-loop quantum gravity test of the ABJM conjecture was done successfully \cite{Bhattacharyya:2012ye}.}
Subsequently, Mari\~no and Putrov developed a more elegant approach, the Fermi gas approach, without making any use of the matrix model techniques or the holomorphic anomaly equations, to compute directly the partition functions of ${\cal N}=3$ and ${\cal N}=2$ CSM theories including the ABJM theory \cite{Marino:2011eh, Marino:2012az}. They found, in particular, a universal Airy function behavior for the ${\cal N}=3$ theories at large $N$ in the small $k$ M-theory regime. These non-planar results were reaffirmed by numerical studies in the case of the ABJM theory \cite{Hanada:2012si}.
Furthermore, the Fermi gas approach was applied to the Wilson loops, exhibiting again the Airy function behavior \cite{Klemm:2012ii}.
Meanwhile, a number of exact computations of the ABJM partition function  were carried out for various values of $N$ and $k$ \cite{Okuyama:2011su, Hatsuda:2012hm, Putrov:2012zi}.
It should also be noted that the nonperturbative effects ${\cal O}(e^{-N})$ of the M- and D-brane type can be systematically studied both in the matrix model \cite{Drukker:2011zy} and the Fermi gas approaches \cite{Marino:2011eh}.

In the unequal rank $N_1\ne N_2$ case of the ABJ theory, the Fermi gas approach thus far has not been applicable, and the study of finite $N_1$ and $N_2$ corrections to the ABJ partition function has not been as much developed as in the ABJM case.  
In this paper, we wish to lay the ground for the study of the ABJ partition function at finite $N_1$ and $N_2$. To this end, we first compute the partition function of the L(2,1) lens space matrix model \cite{Marino:2002fk,Aganagic:2002wv} exactly. By making use of the relation between the lens space and the ABJ matrix models \cite{Marino:2009jd}, we map the lens space partition function to that of the ABJ matrix model by analytically continuing $N_2$ to $-N_2$. With our particular prescription of the analytic continuation, the final answer for the ABJ partition function is given by ${\rm min}(N_1,N_2)$-dimensional integrals and generalizes the ``mirror description'' of the partition function of the ABJM theory \cite{Kapustin:2010xq}. Our result may thus serve as the starting point for the ABJ generalization of the Fermi gas approach.
Meanwhile, we test our prescription against perturbative expansions as well as the Seiberg duality conjecture of \cite{Aharony:2008gk} and find that our final answer perfectly meets the expectations.

The rest of the paper is organized as follows:  In Section \ref{outline_mainresults} we outline our strategy for the calculations of the ABJ partition function and summarize the main result at each pivotal step of the computations. Most of the computational details are relegated to rather extensive appendices. In Section \ref{Examples} we present a few simple examples of our results in order to elucidate otherwise rather complicated general results. In Section \ref{Checks} we state the result of perturbative and nonperturbative checks that we carried out and illustrate with a few simple examples how they were actually done. Section \ref{CandD} is devoted to the conclusions and the discussions.

%%%%%%%%%%%%%%%%%%%%%%%%%%%%%%%%%%%%%%%%%%%%%%%%%%%%%%%%%%

\section{The outline of calculations and main results}
\label{outline_mainresults}

We are going to compute the partition function of the $U(N_1)_k\times U(N_2)_{-k}$ ABJ theory in the matrix model form \cite{Kapustin:2009kz, Hama:2011ea} obtained by the localization technique \cite{Pestun:2007rz}:
\begin{align}
Z_{\rm ABJ}(N_1,N_2)_k={\cal N}_{\rm ABJ}\int\prod_{i=1}^{N_1}{d\mu_i\over 2\pi}\prod_{a=1}^{N_2}{d\nu_a\over 2\pi}
{\Delta_{\rm sh}(\mu)^2\Delta_{\rm sh}(\nu)^2\over\Delta_{\rm ch}(\mu,\nu)^2}
e^{-{1\over 2g_s}\left(\sum_{i=1}^{N_1}\mu_i^2-\sum_{a=1}^{N_2}\nu_a^2\right)}\ ,
\label{ABJMM}
\end{align}
where the $\Delta_{\rm sh}$ factors are the one-loop determinants of the vector multiplets
\begin{align}
\Delta_{\rm sh}(\mu)=\prod_{1\le i<j\le N_1}\left( 2\sinh\left({\mu_i-\mu_j\over 2}\right)\right)\ ,\quad
\Delta_{\rm sh}(\nu)=\prod_{1\le a<b\le N_2}\left( 2\sinh\left({\nu_a-\nu_b\over 2}\right)\right)\ ,
\end{align}
and the $\Delta_{\rm ch}$ factor is the one-loop determinant of the matter multiplets in the bi-fundamental representation
\begin{align}
\Delta_{\rm ch}(\mu,\nu)=\prod_{i=1}^{N_1}\prod_{a=1}^{N_2}\left(2\cosh\left({\mu_i-\nu_a\over 2}\right)\right)\ .
\end{align}
The string coupling $g_s$ is related to the Chern-Simons level $k\in\mathbb{Z}_{\neq 0}$ by
\be
g_s={2\pi i\over k}\ ,
\ee
and 
the factor ${\cal N}_{\rm ABJ}$ in front is the normalization factor \cite{Marino:2011nm}
\be
{\cal N}_{\rm ABJ}:={i^{-{\kappa\over 2}(N_1^2-N_2^2)}\over N_1!N_2!}\ ,\qquad
\kappa:=\sign k\ .
\ee
Note that, because of the relation
\begin{align}
 Z_{\rm ABJ}(N_2,N_1)_{k}
 =Z_{\rm ABJ}(N_1,N_2)_{-k}
 %=Z_{\rm ABJ}(N_1,N_2)_{k}^*
 \ ,
 \label{N1N2swap}
\end{align}
we can assume $N_1\le N_2$ without loss of generality.

%%%%%%%%%%%%%%%%%%%%%%%%%%%%%%%
\subsection{The outline of calculations}

Before going into the details of calculations, we shall first lay out our technical strategy: We adopt the idea employed in the large $N$ analysis of the  ABJ(M) matrix model in \cite{Drukker:2010nc, Marino:2011nm}. Namely, instead of performing the integrals in (\ref{ABJMM}) directly, 

\bigskip
\begin{asparaenum}[(1)]
\item
we first compute the partition of the L(2,1) lens space matrix model \cite{Marino:2002fk,Aganagic:2002wv} 
\begin{align}
\hspace{-.2cm}
Z_{\rm lens}(N_1,N_2)_k={\cal N}_{\rm lens}\int\prod_{i=1}^{N_1}{d\mu_i\over 2\pi}\prod_{a=1}^{N_2}{d\nu_a\over 2\pi}
\Delta_{\rm sh}(\mu)^2\Delta_{\rm sh}(\nu)^2\Delta_{\rm ch}(\mu,\nu)^2
e^{-{1\over 2g_s}\left(\sum_{i=1}^{N_1}\mu_i^2+\sum_{a=1}^{N_2}\nu_a^2\right)}
\label{lensMM}
\end{align}
with the normalization factor 
\be
{\cal N}_{\rm lens}={i^{-{\kappa\over 2}(N_1^2+N_2^2)}\over N_1!N_2!}\ .
\ee

\bigskip
\item
then analytically continue $N_2$ to $-N_2$ to obtain the partition function of ABJ theory \cite{Marino:2009jd}

\be
Z_{\rm ABJ}(N_1,N_2)_k=\lim_{\epsilon\to 0}\,{\cal C}(N_2, \epsilon) Z_{\rm lens}(N_1,-N_2+\epsilon)_k\ ,
\ee
\end{asparaenum}
where the proportionality constant is given in terms of the Barnes $G$-function $G_2(z)$,
\be
{\cal C}(N_2,\epsilon)=(2\pi)^{-N_2}{G_2(N_2+1)\over G_2(-N_2+1+\epsilon)}\ .
\ee

\medskip\noindent
A key observation is that the partition function (\ref{lensMM}) of the lens space matrix model is  \emph{a sum of Gaussian integrals} and can thus be calculated \emph{exactly} in a very elementary manner. The analytic continuation $N_2\to -N_2$, on the other hand, is ambiguous and not as straightforward as one might expect. We find the appropriate prescription for the analytic continuation in two steps: 

\medskip
\begin{asparaenum}[(2.i)]
\item In the first step we propose a natural prescription that correctly
reproduces, after a generalized $\zeta$-function regularization, the
known perturbative expansions in the string coupling $g_s$. The
resulting expression, however, is a formal series that is non-convergent
and singular when $k$ is an even integer.

\medskip
\item To circumvent these issues, in the second step, we introduce an integral representation which renders a formal series perfectly well-defined. 
\end{asparaenum}

\medskip\noindent
In other words, the integral representation (A) implements a generalized $\zeta$-function regularization automatically and (B) provides an analytic continuation in the complex parameter $g_s$ for the formal series.

\medskip As we will see later, the final answer in the integral
representation passes perturbative as well as some nonperturbative tests
and generalizes the ``mirror description'' \cite{Kapustin:2010xq}
of the partition function of the ABJM theory to the ABJ theory.

%%%%%%%%%%%%%%%%%%%%%%%%%%%%%%%%%%%%%%%%%%%%%%%%%%%%%%%%%%

\subsection{The main results}
\label{mainresults}

We present, without much detail of derivations, the main result at each step of the outlined calculations. Most of the technical details will be given in the appendices.

%%%%%%%%%%%%%%%%%%%%%%%%%%%%%%%%%%%%%%%%%%%%%%%%%%%%%%%%%%

\subsubsection*{$\bullet$ The lens space matrix model}

As emphasized above, the lens space partition function (\ref{lensMM}) is a sum of Gaussian integrals and can be calculated exactly:  
\begin{align}
 Z_{\text{lens}}(N_1,N_2)_k
 &=
 i^{-{\kappa\over 2}(N_1^2+N_2^2)}
 \left({g_s\over 2\pi}\right)^{N\over 2}
 q^{-{1\over 3}N(N^2-1)}
 \notag\\
 &\qquad
 \times \sum_{(\cN_1,\,\cN_2)}
 \prod_{C_j<C_k} (q^{C_j}-q^{C_k})
 \prod_{D_a<D_b} (q^{D_a}-q^{D_b})
 \prod_{C_j,D_a} (q^{C_j}+q^{D_a})\ ,\label{lensAns1}
\end{align}
where
\begin{align}
 q:=e^{-g_s}=e^{-{2\pi i \over k}}, \qquad N=N_1+N_2.\label{qgs}
\end{align}
The symbol $(\cN_1,\,\cN_2)$ denotes the partition of the numbers $(1, 2, \cdots, N)$ into two groups $\cN_1=(C_1, C_2, \cdots, C_{N_1})$ and $\cN_2=(D_1, D_2, \cdots, D_{N_2})$ where $C_i$'s and $D_a$'s are ordered as $C_1<\cdots<C_{N_1}$ and $D_1<\cdots<D_{N_2}$.
The computation proceeds in two steps: (1) Gaussian integrals and (2) sums over permutations. 
The detailed derivation can be found in Appendix \ref{Appendix_LensMM}\@.

As advertised, the result (\ref{lensAns1}) can be written as a product
of $q$-deformed Barnes $G$-function and a generalization of multiple
$q$-hypergeometric function:
\begin{align}
\hspace{-.4cm}
 Z_{\text{lens}}(N_1,N_2)_k
 &=
 i^{-{\kappa \over 2}(N_1^2+N_2^2)}
 \left({g_s\over 2\pi}\right)^{N\over 2}
 q^{-{1\over 6}N(N^2-1)}(1-q)^{\half N(N-1)}G_2(N+1;q)\,S(N_1,N_2)\ ,
 \label{lensAns2}
\end{align}
where
\begin{align}
 S(N_1,N_2)= \sum_{(\cN_1,\,\cN_2)}
 \prod\limits_{C_j<D_a}{q^{C_j}+q^{D_a}\over q^{C_j}-q^{D_a}}
 \prod\limits_{D_a<C_j}{q^{D_a}+q^{C_j}\over q^{D_a}-q^{C_j}}\ .
 \label{Sfunction}
\end{align}
The $q$-deformed Barnes $G$-function $G_2(z;q)$ is defined in Appendix
\ref{Appendix_q-Analogs} and, as will be elaborated later, $S(N_1,
N_2)$ is a generalization of multiple $q$-hypergeometric function.
Recalling that $q=e^{-g_s}$, it is rather fascinating to observe that
the string coupling $g_s$ is not only the loop-expansion parameter in
quantum mechanics but also a quantum deformation parameter of special
functions.

In Section \ref{Examples} we will give simple examples of the lens space partition function in order to elucidate the $q$-hypergeometric structure.

%%%%%%%%%%%%%%%%%%%%%%%%%%%%%%%%%%%%%%%%%%%%%%%%%%%%%%%%%%

\subsubsection*{$\bullet$ The ABJ theory}

The next step in our strategy is the analytic continuation $N_2\to -N_2$ which maps the partition function of the lens space matrix model to that of the ABJ theory. For this purpose, we find it convenient to work with the second expression of the lens space partition function (\ref{lensAns2}).  Our claim is that the analytic continuation yields the following expression for the ABJ partition function in a formal series 
\begin{align}
Z_{\rm ABJ}(N_1, N_2)_k&= i^{-{\kappa \over 2}(N_1^2+N_2^2)}
(-1)^{\half N_1(N_1-1)}\, 2^{-N_1}\,
\left({g_s\over 2\pi}\right)^{N_1+N_2\over 2} (1-q)^{M(M-1)\over 2}
G_2(M+1;q)\nn\\
&\quad\times
  {1\over N_1!}\sum\limits_{s_1,\dots,s_{N_1}\ge 0}
  (-1)^{s_1+\cdots+s_{N_1}}
  \prod\limits_{j=1}^{N_1}{(q^{s_j+1})_{M}\over (-q^{s_j+1})_{M}}
  \prod\limits_{j<k}^{N_1}
  {(1-q^{s_k-s_j})^2\over (1+q^{s_k-s_j})^2}\ .
\label{ABJpart_formalsum}
\end{align}
where we defined $M=N_2-N_1$ (for $N_2>N_1)$ and $(a)_n$ is a shorthand
notation for the $q$-Pochhammer symbol $(a;q)_n$ defined in Appendix
\ref{Appendix_q-Analogs}\@. We used an $\epsilon$-prescription in
continuing $N_2$ to $-N_2$, as explained in detail in Appendix
\ref{Appendix_AC}\@.

However, as noted above, there are in principle multiple ways to continue $N_2$ to $-N_2$. It thus requires a particular prescription to fix this ambiguity. Our prescription is to continue $N_2$ to $-N_2$ with $S(N_1, N_2)$ written in the form
\begin{align}
S(N_1, N_2)=\gamma(N_1, N_2)\Psi(N_1, N_2)
\label{gammaPsi}
\end{align}
where
\begin{align}
\gamma(N_1,N_2)&=
 (-1)^{\half N_1(N_1-1)}
 \prod_{j=1}^{N_1-1}{(-q)_j^2\over (q)_j^2}
 \prod_{j=1}^{N_1}
  {(-q^j)_{N_2}(-q^j)_{-N_1-N_2}\over (q^j)_{N_2}(q^j)_{-N_1-N_2}}\ ,\label{gamma}
 \\
 \Psi(N_1,N_2)&=
 {1\over N_1!}\sum_{s_1,\dots,s_{N_1}\ge 0}
 (-1)^{s_1+\cdots+s_{N_1}}
 \prod_{j=1}^{N_1}{(q^{s_j+1})_{-N_1-N_2}\over (-q^{s_j+1})_{-N_1-N_2}}
 \prod_{1\le j<k\le N_1}
 {(q^{s_k-s_j})_1^2\over (-q^{s_k-s_j})_1^2}\ .
  \label{Psi}
\end{align}
As will be explained more in detail in Appendix \ref{Appendix_AnalCont},
there are a number of ways to express $S(N_1, N_2)$ that could yield
different results after the analytic continuation: The range of the sum in
(\ref{Sfunction}) runs from $1$ to $N=N_1+N_2$. In order to make sense
of analytic continuation in $N_2 (> N_1)$, the finite sum
(\ref{Sfunction}) is extended to the infinite sum (\ref{Psi}). In
fact, the summand for $s_i>N-1$ in (\ref{gammaPsi}) vanishes after an
appropriate regularization. Now the point is that these vanishing terms
could yield nonvanishing contributions after the analytic
continuation. Clearly, the way to extend the finite sums to infinite
ones is not unique, and this is where the ambiguity lies.

Our guideline for the correct prescription is to successfully reproduce the perturbative expansions in $g_s$. Indeed, it can be checked that the formal series (\ref{ABJpart_formalsum}) has the correct perturbative expansions, as we will discuss more in Section \ref{PE}.

%%%%%%%%%%%%%%%%%%%%%%%%%%%%%%%%%%%%%%%%%%%%%%%%%%%%%%%%%%
\subsubsection*{$\bullet$ The integral representation}

As alluded to in the outline, the result (\ref{ABJpart_formalsum}) is
not the final answer. It is a formal series that is non-convergent and
singular when $k$ is an even integer. It can be rendered perfectly
well-defined by introducing an integral representation: Specifically,
our final answer for the analytic continuation is
\begin{align}
Z_{\rm ABJ}(N_1, N_2)_k&=i^{-{\kappa \over 2}(N_1^2+N_2^2)}
 (-1)^{\half N_1(N_1-1)}\, 2^{-N_1}\,
\left({g_s\over 2\pi}\right)^{N_1+N_2\over 2}  (1-q)^{M(M-1)\over 2}
G_2(M+1;q)\nn\\
&\quad\times
  {1\over N_1!}\prod_{j=1}^{N_1}\left[{-1\over 2\pi i}
  \int_{I} {\pi \, ds_j\over \sin(\pi s_j)}\right]
 \prod_{j=1}^{N_1}{(q^{s_j+1})_{M}\over (-q^{s_j+1})_{M}}
 \prod_{1\le j<k\le N_1}
 {(1-q^{s_k-s_j})^2\over (1+q^{s_k-s_j})^2}\ ,
\label{ABJpart_integral}
\end{align}
where $M=N_2-N_1$ (for $N_2>N_1$) and the integration range $I=\left[-i\infty-\eta, +i \infty-\eta\right]$ with $\eta>0$. We note that there is a subtlety in the choice of $\eta$: For example, when the string coupling $g_s$ takes the actual value of our interest, ${2\pi i\over k}$ with an integer $k$,  as we will elaborate in Section \ref{SD}, the parameter $\eta$ should be varied so that the partition function remains analytic in $k$, as one decreases the value of $k$ from the small coupling regime $|g_s|=|2\pi i/k|\ll 1$. 

\begin{figure}[h!]
\centering \includegraphics[height=2.5in]{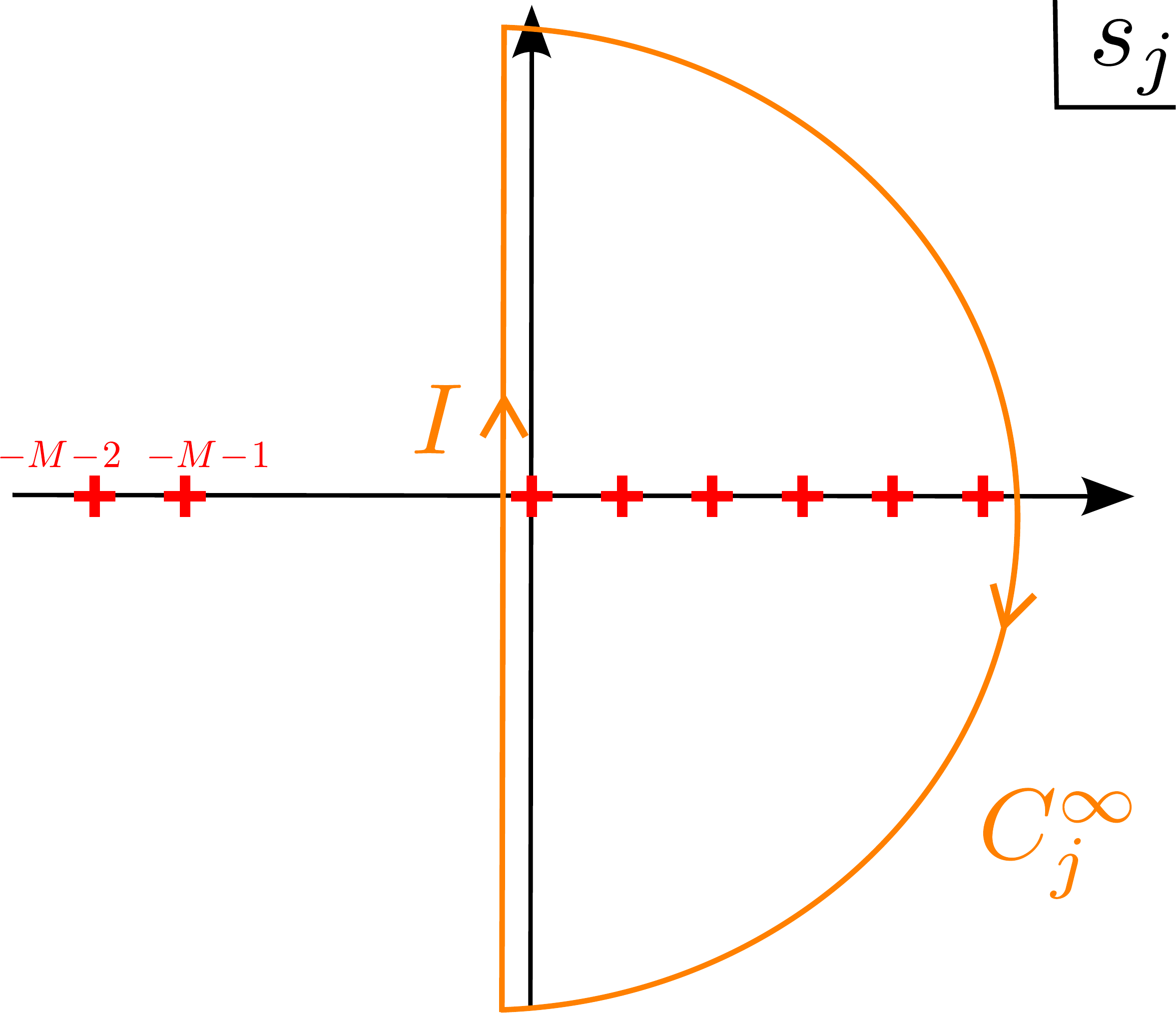} \caption{\sl ``The
integration contour'' $C_j=I+C_j^{\infty}$ for the perturbative ABJ
partition function: The only perturbative (P) poles are indicated by
red ``$+$''. See text for detail.}  \label{contour1}
\end{figure}

Although we lack a first principle derivation of the integral
representation, we can give heuristic arguments as follows:
First, this integral representation ``agrees'' with the formal series
(\ref{ABJpart_formalsum}) order by order in the perturbative
$g_s$-expansions. The integrals could be evaluated by considering the
closed contours $C_j$ composed of the vertical line $I$ and the
infinitely large semi-circle $C_j^{\infty}$ on the right half of the
complex $s_j$-plane, if the contribution from $C_j^{\infty}$ were to
vanish; see Figure \ref{contour1}.  In the $g_s$-expansions, the poles
would only come from the factors $1/\sin(\pi s_j)$ and are at
$s_j=n_j\in \mathbb{Z}_{\ge 0}$.  Thus the residue integrals would
correctly reproduce (\ref{ABJpart_formalsum}).  In actuality, however, the
contribution from $C_j^{\infty}$ does not vanish, and thus this argument is
heuristic at best; we will see precisely how the $g_s$-expansions work in
an example in section \ref{PE}.  We note that, to the same degree of imprecision, the integral representation (\ref{ABJpart_integral}) can be
thought of as the Sommerfeld-Watson transform of
\eqref{ABJpart_formalsum}.\footnote{We thank Yoichi Kazama and Tamiaki
Yoneya for pointing this out to us.}

\begin{figure}[h!]
\centering \includegraphics[height=2.2in]{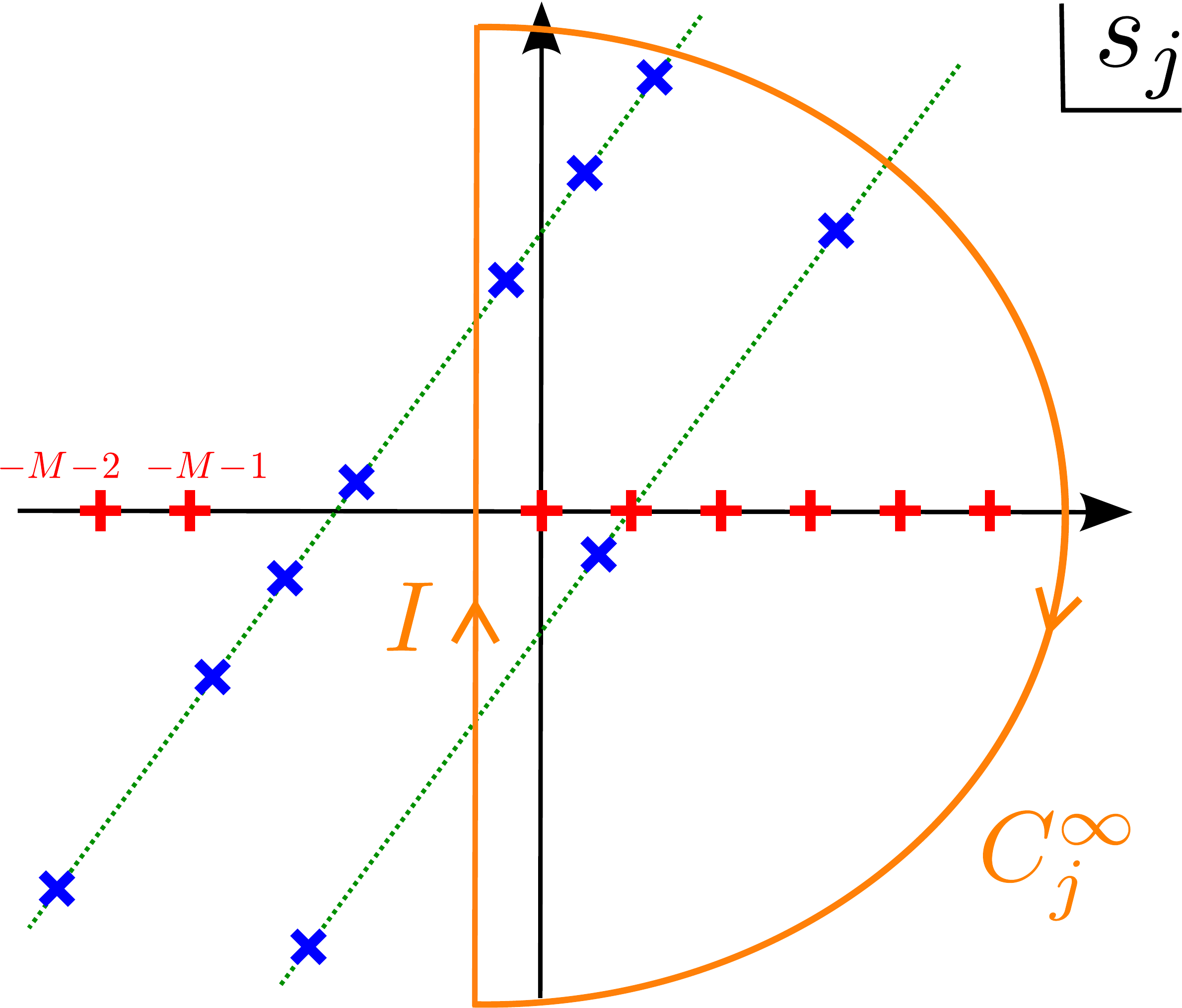} \hspace{1cm}
\includegraphics[height=2.2in]{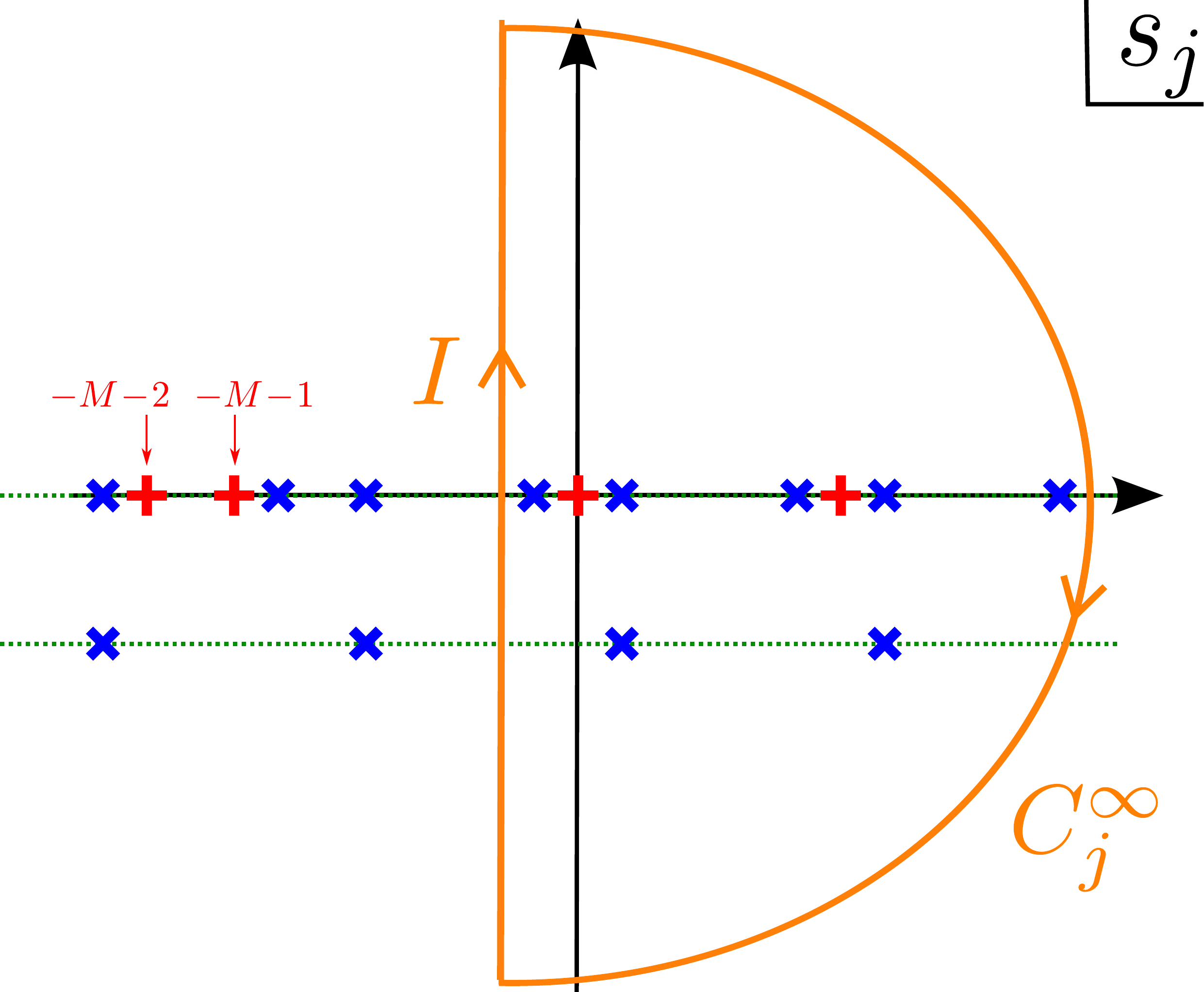} \hspace{.45cm} \caption{\sl The
nonperturbative (NP) poles are added and indicated by blue
``$\times$''. The left panel corresponds to the complex $g_s$ case. The
right panel is the actual case of our interest $g_s=2\pi i/k$. (Shown is
the case $k=3$ and $M=3$.)}  \label{NPpoles}
\end{figure}

Second, as implied in the first point, the integral representation
(\ref{ABJpart_integral}) provides a ``nonperturbative completion''
for the formal series (\ref{ABJpart_formalsum}).  In fact,
nonperturbatively, there appear additional poles from the factors
$1/(-q^{s_j+1})_M$ and $1/(1+q^{s_k-s_j})^2$ in the contour
integrals. They are located at $s_j=-{(2n+1)\pi i\over g_s}-m$ and
$s_j=-{(2n+1)\pi i\over g_s}+s_k$ with $n\in\mathbb{Z}$ and $m=1,\cdots,
M$, as shown in Figure \ref{NPpoles}. Their residues are of order
$e^{1/g_s}$. Hence these can be regarded as
nonperturbative (NP) poles, whereas the previous ones are perturbative
(P) poles. Again, these statements are rather heuristic, and we will see how precisely P and NP
poles contribute to the contour integral in Section \ref{SD}.

A few remarks are in order: 

\begin{asparaenum}[(1)]

\medskip
\item As promised, there is no issue of convergence in the expression
(\ref{ABJpart_integral}). It is also well-defined in the entire complex
$q$-plane. The integrand becomes singular for $q=e^{-2\pi i/k}$ with
even integer $k$ as in the formal series
(\ref{ABJpart_formalsum}). However, this merely represents pole
singularities and yields finite residue contributions.

\medskip
\item
It should be noted that our main result (\ref{ABJpart_integral}) lacks a first principle derivation. It thus requires {\it a posteriori\/} justification. On this score, as stressed and will be discussed more in Section \ref{PE}, the integral representation (\ref{ABJpart_integral}) correctly reproduces the perturbative expansions; moreover, it automatically implements a generalized $\zeta$-function regularization needed in the perturbative expansions of the infinite sum (\ref{ABJpart_formalsum}). Meanwhile, a successful test of the Seiberg duality conjectured in \cite{Aharony:2008gk} provides evidence for our proposed nonperturbative completion. 
We will explicitly show a few nontrivial examples of the Seiberg duality at work in Section \ref{SD}.

\medskip
\item
In the ABJM limit ($M=0$), the integral representation
(\ref{ABJpart_integral}) coincides with the ``mirror description'' of
the ABJM partition function found in \cite{Kapustin:2010xq}. This
provides a further support for our prescription and implies that we have
found a generalization of the ``mirror description'' in the case of the ABJ
theory. Our finding may thus serve as the starting point for the
generalization of the Fermi gas approach developed in
\cite{Marino:2011eh} to the ABJ theory.

\medskip
\item
One of the ABJ  conjectures is that the ${\cal N}=6$ $U(N_1)_k\times U(N_1+M)_{-k}$ theory with $M>k$ may not exist as a unitary theory \cite{Aharony:2008gk}. It is further expected that the supersymmetries are spontaneously broken in this case \cite{Bergman:1999na} (see also 
\cite{Hashimoto:2010bq}). A manifestation of this conjecture is that the partition function (\ref{ABJpart_integral}) vanishes when $M > k$ because
\be
(1-q)^{{M(M-1)\over 2}}G_2(M+1;q)=\prod_{j=1}^{M-1}(q)_j=0\quad
\mbox{for}\quad q=e^{-{2\pi i\over k}}\ .
\ee
Note that the $q$-deformed Barnes $G$-function $G_2(M+1;q)$ is precisely a factor that appears in the partition function of the $U(M)_k$ Chern-Simons theory. We thus expect that this property is not peculiar to the ${\cal N}=6$ CSM theories but holds for CSM theories with less supersymmetries as long as they contain the $U(M)_k$ CS theory as a subsector.\footnote{We thank Vasilis Niarchos for discussions on this point.}

\end{asparaenum}

%%%%%%%%%%%%%%%%%%%%%%%%%%%%%%%%%%%%%%%%%%%%%%%%%%%%%%%%%%

\section{Examples}
\label{Examples}

In this section we present a few simple examples of the lens space and ABJ partition functions in order to get the feel of the expressions found in the previous section. In particular, these examples clarify the appearance of $q$-hypergeometric functions in the lens space partition function and how they are mapped to in the ABJ partition function. We also provide a simplest example of the exact ABJ partition function.

%%%%%%%%%%%%%%%%%%%%%%%%%%%%%%%%%%%%%%%%%%%%%%%%%%%%%%%%%%
\subsubsection*{$\bullet$ The CS matrix model}

The first example is the simplest case, the $N_1=0$ or $N_2=0$ case, which corresponds to the Chern-Simons matrix model. From (\ref{lensAns2}) one immediately finds for the $U(M)_k$ CS theory that
\be
Z_{\rm CS}(M)_k=Z_{\rm lens}(M,0)_k= i^{-{\kappa M(M-1)\over 2}}
 |k|^{-{M\over 2}}
 q^{-{M(M^2-1)\over 6}}(1-q)^{\half M(M-1)}G_2(M+1;q)\ .
\label{CSMM}
\ee
Note that this takes the more familiar form \cite{Marino:2004uf, Tierz:2002jj} (without the level shift) if one uses the formula
\be
i^{-{\kappa M(M-1)\over 2}}(1-q)^{\half M(M-1)}G_2(M+1;q)=q^{{M(M^2-1)\over 12}}\prod_{j=1}^{M-1}\left(2\sin{\pi j\over |k|}\right)^{M-j}\ .
\ee
It should now be clear that the $q$-deformed Barnes $G$-function is a contribution from the $U(|N_1-N_2|)_k$ pure CS subsector in the $U(N_1)_k\times U(N_2)_{-k}$ theory. 

%%%%%%%%%%%%%%%%%%%%%%%%%%%%%%%%%%%%%%%%%%%%%%%%%%%%%%%%%%
\subsubsection*{$\bullet$ The lens space matrix model}

The next simplest example is the $N_1=1$ case studied in detail in Appendix \ref{Appendix_N1equalto1}\@. From (\ref{lensAns2}) together with (\ref{mckp3Sep12}) and (\ref{evuj11Sep12}), the $U(1)_k\times U(N_2)_{-k}$ lens space partition function yields
\begin{align}
Z_{\text{lens}}(1,N_2)_k
 = \,&
 i^{-{\kappa\over 2}(N_2^2+1)}
 \left({g_s\over 2\pi}\right)^{N_2+1\over 2}
 q^{-{N_2(N_2+1)(N_2+2)\over 6}}(1-q)^{N_2(N_2+1)\over 2}G_2(N_2+2;q)\nn\\
&\times {(-q)_{N_2}\over (q)_{N_2}}\, _2\phi_1\left({q^{-N_2},-q\atop -q^{-N_2}} ; q,-1\right)\ ,
\label{lens1N2}
\end{align}
where the special function $\, _2\phi_1=\Phi(1,N_2)$ is a
$q$-hypergeometric function \cite{Gasper-Rahman} whose definition is given in Appendix \ref{Appendix_q-Analogs}\@. Intriguingly, the
whole function $S(1,N_2)$ in the second line is essentially an
orthogonal $q$-polynomial, the continuous $q$-ultraspherical (or Rogers)
polynomial \cite{Koekoek-Swarttouw}, and very closely related to Schur
Q-polynomials \cite{Rosengren:2006}.

\medskip
The next example is the $N_1=2$ case discussed in detail in Appendix \ref{Appendix_N1equalto2}\@. In parallel with the previous case, from (\ref{lensAns2}) together with (\ref{kgmb7Nov12}) and (\ref{gyiq5Sep12}), one finds the $U(2)_k\times U(N_2)_{-k}$ lens space partition function
\begin{align}
\hspace{-.3cm}
Z_{\text{lens}}(2,N_2)_k
 &= 
 i^{-{\kappa\over 2}(N_2^2+4)}
 \left({g_s\over 2\pi}\right)^{N_2+2\over 2}
 q^{-{(N_2+1)(N_2+2)(N_2+3)\over 6}}(1-q)^{(N_2+1)(N_2+2)\over 2}G_2(N_2+3;q)\nn\\
&\times {(-q)_{N_2}(-q^2)_{N_2}\over (q)_{N_2}(q^2)_{N_2}}\,
\Phi^{2:2;4}_{2:1;3}\!\left(
 \begin{array}{c@{~}c@{~}c@{~}c@{~}c}
  q^{-N_2},-q^2&:&q^{-N_2-1},-q&;& q^2,q^2, -q,-q \\
  -q^{-N_2},q^2&:& -q^{-N_2-1} &;&  -q^2,-q^2, q
 \end{array}
 ;~q ~;1,-1\right)\ ,
\label{lens2N2}
\end{align}
where the special function $\Phi^{2:2;4}_{2:1;3}=\Phi(2,N_2)$ is a double $q$-hypergeometric function defined in Section 10.2 of \cite{Gasper-Rahman}.

\medskip
As promised, these examples elucidate that the function $S(N_1,N_2)$ defined in (\ref{Sfunction}) is a generalization of multiple $q$-hypergeometric function.

%%%%%%%%%%%%%%%%%%%%%%%%%%%%%%%%%%%%%%%%%%%%%%%%%%%%%%%%%%
\subsubsection*{$\bullet$ The ABJ theory}

We now present the ABJ counterpart of the previous two examples. Although we have placed great emphasis on the $q$-hypergeometric structure of the lens space partition function, we have not found a way to take full advantage of this fact in understanding the ABJ partition function thus far. 

In the meantime, as mentioned in the previous section and discussed in great detail in Appendix \ref{Appendix_AC}, we find the expression (\ref{moge23Sep12}) more convenient for performing the analytic continuation $N_2\to -N_2$ than the $q$-hypergeometric representation (\ref{fvrs7Sep12}). The end result is presented in (\ref{ABJpart_integral}). In the case of the $U(1)_k\times U(N_2)_{-k}$ ABJ partition function, one finds
%\begin{align}
%Z_{\rm ABJ}(1, N_2)_k=\,&{1\over 2} \,i^{-{\kappa\over 2}(N_2-1)(N_2-2)}\,
%|k|^{-{N_2+1\over 2}}(1-q)^{(N_2-1)(N_2-2)\over 2}\,
%G_2(N_2;q)\nn\\
%&\hspace{0.1cm}\times \left[{-1\over 2\pi i}
%  \int_{I} {\pi \, ds\over \sin(\pi s)}
%\prod_{l=1}^{N_2-1}\tan\left({(s+l)\pi\over |k|}\right)\right]\ .
%\label{ABJN11}
%\end{align}
\begin{align}
Z_{\rm ABJ}(1, N_2)_k=\,&{1\over 2} \,q^{{1\over 12}N_2(N_2-1)(N_2-2)}
|k|^{-{N_2+1\over 2}}\prod_{j=1}^{N_2-2}\left(2\sin{\pi j\over |k|}\right)^{N_2-1-j}
\nn\\
&\hspace{0.1cm}\times \left[{-1\over 2\pi i}
  \int_{I} {\pi \, ds\over \sin(\pi s)}
\prod_{l=1}^{N_2-1}\tan\left({(s+l)\pi\over |k|}\right)\right]\ .
\label{ABJN11}
\end{align}
Similarly, the $U(2)_k\times U(N_2)_{-k}$ ABJ partition function yields
%\begin{align}
%Z_{\rm ABJ}(2, N_2)_k=&{i\over 8}(-1)^{N_2-1}i^{-{\kappa\over 2}N_2(N_2-1)}
%|k|^{-{N_2+2\over 2}}  (1-q)^{(N_2-2)(N_2-3)\over 2}
%G_2(N_2-1;q)\nn\\
%&\times
% \prod_{j=1}^{2}\left[{-1\over 2\pi i}
%  \int_{I} {\pi \, ds_j\over \sin(\pi s_j)}
%\prod_{l=1}^{N_2-2}\tan\left({(s_j+l)\pi\over |k|}\right)\right]
%\tan^2\left({(s_2-s_1)\pi \over |k|}\right)\ .
%\label{ABJN12}
%\end{align}
\begin{align}
Z_{\rm ABJ}(2, N_2)_k
 =\,&
 -{1\over 8}q^{{1\over 12}(N_2-1)(N_2-2)(N_2-3)}
|k|^{-{N_2+2\over 2}}\prod_{j=1}^{N_2-3}\left(2\sin{\pi j\over |k|}\right)^{N_2-2-j}
 \nn\\
&\times
 \prod_{j=1}^{2}\left[{-1\over 2\pi i}
  \int_{I} {\pi \, ds_j\over \sin(\pi s_j)}
\prod_{l=1}^{N_2-2}\tan\left({(s_j+l)\pi\over |k|}\right)\right]
\tan^2\left({(s_2-s_1)\pi \over |k|}\right)\ .
\label{ABJN12}
\end{align}
Note that the $U(N_1)_k\times U(N_2)_{-k}$ ABJ theory with finite $N_1$
and large $N_2$ and $k$ is conjectured to be dual to ${\cal N}=6$
parity-violating Vasiliev higher spin theory on $AdS_4$ with $U(N_1)$
gauge symmetry \cite{Giombi:2011kc, Chang:2012kt}.  It would thus be
very interesting to study the large $N_2$ and $k$ limit of the $N_1=1$
and $2$ partition functions \cite{WIP1}. It may shed some lights on the
understanding of the ${\cal N}=6$ parity-violating Vasiliev theory on
$AdS_4$.\footnote{Since higher spin theories are inherently dual to
vector models \cite{Sundborg:2000wp, Sezgin:2002rt, Klebanov:2002ja},
the ABJ theory apparently contains more degrees of freedom than higher
spin fields \cite{Aharony:2011jz}. Those extra degrees of freedom are
the large $N_2$ dual of the $U(N_2)$ Chern-Simons theory and thus topological closed
strings \cite{Gopakumar:1998ki}. It is then plausible to expect that the
higher spin partition function is given by the ratio $Z_{\rm ABJ}/Z_{\rm
CS}$. We thank Hiroyuki Fuji and Xi Yin for related discussions.}

\medskip
Finally, we provide a simplest example of the exact ABJ partition function, {\it i.e.}, the $U(1)_k\times U(2)_{-k}$ case. The integral in (\ref{ABJN11}) can be carried out by applying a similar trick to the one used in \cite{Okuyama:2011su}. This yields
\begin{align}
Z_{\rm ABJ}(1, 2)_k=\,&{1\over 2} \,
|k|^{-{3\over 2}}\times
\begin{cases}
\half\left[\sum_{n=1}^{|k|-1}(-1)^{n-1}\tan\bigl({\pi n\over |k|}\bigr)+|k|(-1)^{|k|-1\over 2}\right] & \quad(k={\rm odd})\ ,\\[1ex]
\sum_{n=1}^{|k|-1}(-1)^{n-1}\left({1\over 2}-{n\over k}\right)\tan\bigl({\pi n\over |k|}\bigr) & \quad(k={\rm even})\ .
\end{cases}
\end{align}
%\begin{align}
%Z_{\rm ABJ}(1, 2)_k=\,&-{1\over 4} \,i^{7\over 2}\,
%k^{-{3\over 2}}\left\{
%\begin{array}{ll}
%\left(\sum_{n=1}^{k-1}{(-1)^n\over 1+e^{2\pi in\over k}}+ {i(-1)^{k-1\over 2}k\over 2}\right) & \quad(k={\rm odd})\ ,\\
%\sum_{n=1}^{k-1}{(-1)^n\left(1-{2n\over k}\right)\over 1+e^{2\pi in\over k}} & \quad(k={\rm even})\ .
%\end{array}
%\right.
%\end{align}
It may be worth noting that the formal series (\ref{ABJpart_formalsum}) for the $U(1)_k\times U(2)_{-k}$ theory, albeit nonconvergent, can be expressed in a closed form after a regularization:  
\begin{align}
Z_{\rm ABJ}(1, 2)_k=\,&{1\over 2} \,i^{-\kappa}\,
|k|^{-{3\over 2}}\left[\half -{2\over\log q}\left(\log\left(1+q^2\over 1+q\right)+\psi_q(1)-2\psi_{q^2}(1)+\psi_{q^4}(1)\right)\right]\ ,
\end{align}
where $\psi_q(z)$ is a $q$-digamma function defined in Appendix
\ref{Appendix_q-Analogs}, and we used the regularization
$\sum_{s=0}^{\infty}(-1)^s=\half$. This expression is, however, not
well-defined for $q$ a root of unity and hence an integer $k$. On the
other hand, this exemplifies the fact that the integral representation
(\ref{ABJpart_integral}) provides an analytic continuation of the formal
series (\ref{ABJpart_formalsum}) in the complex $q$-plane.

%%%%%%%%%%%%%%%%%%%%%%%%%%%%%%%%%%%%%%%%%%%%%%%%%%%%%%%%%%

\section{Checks}
\label{Checks}

As mentioned in Section \ref{outline_mainresults}, our main result
(\ref{ABJpart_integral}) lacks a first principle derivation. It thus
requires {\it a posteriori\/} justification. In this section we show that our
prescription passes perturbative as well as nonperturbative tests. We
have, however, been unable to prove it in generality. Although our
checks are on a case-by-case basis, we have examined several nontrivial
cases that provide convincing evidence for our claim.%
\footnote{
We also recall that, in the ABJM case $N_1=N_2$, the expression
(\ref{ABJpart_integral}) reproduces the ``mirror description'' of
the ABJM partition function \cite{Kapustin:2010xq}. Furthermore, for
simple cases such as $(N_1,N_2)=(1,2),(1,3)$, it is possible to
explicitly carry out the ABJ matrix integral \eqref{ABJMM} and check
that it agrees with the expression (\ref{ABJpart_integral}) for all $k$.
}

%%%%%%%%%%%%%%%%%%%%%%%%%%%%%%%%%
\subsection{Perturbative expansions}
\label{PE}

The perturbative expansion of the lens space free energy is presented in \cite{Aganagic:2002wv}.  In Appendix \ref{Appendix_PE}, we extend their result to the order ${\cal O}(g_s^8)$. 
We would like to see if the perturbative expansions of both (\ref{ABJpart_formalsum}) and (\ref{ABJpart_integral}) correctly reproduce this result with the replacement $N_2$ by $-N_2$. We have checked the cases $N_1=1$ and $N_2$ up to $8$, $N_1=2$ and $N_2$ up to $5$, $N_1=3$ and $N_2$ up to $5$, and $N_1=4$ and $N_2$ up to $4$, to the order ${\cal O}(g_s^8)$ and found perfect agreements with the result in Appendix \ref{Appendix_PE}\@. These checks are straightforward, and we will not spell out all the details. Instead, we describe only the essential points in the calculations and illustrate with a simple but nontrivial example how the checks were done in detail.

%%%%%%%%%%%%%%%%%%%%%%%%%%
\subsubsection*{The formal series}

In the case of the formal series (\ref{ABJpart_formalsum}), as remarked in the previous section, the perturbative expansion is correctly reproduced after the generalized $\zeta$-function regularization:
\be
\sum_{s=0}^{\infty}(-1)^ss^n=\left\{
\begin{array}{ll}
{\rm Li}_{-n}(-1)=(2^{n+1}-1)\zeta(-n)
=-{2^{n+1}-1\over n+1}B_{n+1} &\quad(\mbox{for}\quad n\ge 1)\ ,\\
1+{\rm Li}_0(-1) = \half=-B_1 &\quad(\mbox{for}\quad n=0)\ ,
\end{array}
\right.
\label{genZetareg}
\ee
where ${\rm Li}_s(z)$ is the polylogarithm and $B_n$ are the Bernoulli numbers.
We show the detail of the $(N_1,N_2)=(2,3)$ example to illustrate how the generalized $\zeta$-function regularization yields the correct perturbative expansion to the order ${\cal O}(g_s^4)$. In this case there are two infinite sums involved. Now, recall that the summand is a function of $q=\exp\left(-g_s\right)$. Expanding it as a power series in $g_s$ and using the  regularization (\ref{genZetareg}), one finds 
\begin{align}
\hspace{-.2cm}
\text{The 2nd line of (\ref{ABJpart_formalsum})}&={g_s^4\over 32}\left(
\text{Li}^2_{-3,-1}\!-\!2 \text{Li}^2_{-2,-2}\!+\!\text{Li}^2_{-1,-3}\right)\!-\!{g_s^6\over 384}\left(3 \text{Li}^2_{-5,-1}\!-\!10 \text{Li}^2_{-4,-2}\!+\!14 \text{Li}^2_{-3,-3}\right.\nn\\
&\left.-10 \text{Li}^2_{-2,-4}+3 \text{Li}^2_{-1,-5}\right)
+{g_s^8\over 23040}\left(33 \text{Li}^2_{-7,-1}-154 \text{Li}^2_{-6,-2}+336 \text{Li}^2_{-5,-3}\right.\nn\\
&\left.-430 \text{Li}^2_{-4,-4}+336 \text{Li}^2_{-3,-5}-154 \text{Li}^2_{-2,-6}+33 \text{Li}^2_{-1,-7}\right)+{\cal O}(g_s^{10})\nn\\
&=-\frac{1}{512}g_s^4-\frac{19}{12288}g_s^6-\frac{137}{81920}g_s^8+{\cal O}(g_s^{10})\ ,
\end{align}
where we abbreviated the product ${\rm Li}_{-n_1}(-1){\rm Li}_{-n_2}(-1)$ by ${\rm Li}^2_{-n_1,-n_2}$. This yields
\be
F_{\rm ABJ}(2,3)=\log Z_{\rm ABJ}(2,3)=
\log\left[\frac{2^{-12}(2\pi g_s)^{13\over 2}}{ 2^{-1}(2\pi)^{9}}\right]
+\frac{19}{24}g_s^2+\frac{3127}{5760}g_s^4+{\cal O}(g_s^6)
\ee
in agreement with the result in Appendix \ref{Appendix_PE} with the replacement $N_2$ by $-N_2$. Note also that the tree contribution, the first logarithmic term, is in a precise agreement with (\ref{nimf2Nov12}).

%%%%%%%%%%%%%%%%%%%%%%%%%%
\subsubsection*{The integral representation}

The integral representation (\ref{ABJpart_integral}) does not require any regularization. Instead, the generalized $\zeta$-function regularization (\ref{genZetareg}) is automatically implemented by the integral
\be
-{1\over 2\pi i}\int_I{\pi ds\over\sin(\pi s)}s^n =-{2^{n+1}-1\over n+1}B_{n+1}\ ,
\ee
where $n\ge 0$. It follows immediately from this fact that 
\be
\text{the 2nd line of (\ref{ABJpart_integral})}=\text{the 2nd line of (\ref{ABJpart_formalsum})}
\ee
at all orders in the $g_s$-expansions. Hence the integral representation correctly reproduces the perturbative expansions.

%%%%%%%%%%%%%%%%%%%%%%%%%%%%%%%%%
\subsection{The Seiberg duality}
\label{SD}

As emphasized before, the integral representation
(\ref{ABJpart_integral}) provides a ``nonperturbative completion''
for the formal series (\ref{ABJpart_formalsum}). A way to test this
claim is to see if the Seiberg duality conjectured in
\cite{Aharony:2008gk} holds.\footnote{This duality is a special case of
the Giveon-Kutasov duality of ${\cal N}=2$ CS theories
\cite{Giveon:2008zn} that is further generalized to theories with
fundamental and adjoint matter by Niarchos \cite{Niarchos:2008jb}.} This
duality is an equivalence between the two ABJ theories; schematically,
\be
U(N_1)_k\times U(N_1+M)_{-k} = U(N_1+|k|-M)_k\times U(N_1)_{-k}\ .
\label{Seibergduality}
\ee
We are going to show, in the simple but nontrivial case of 
$N_1=1$, that the partition functions of the dual pairs agree up to a
phase.  In fact, a proof of the Giveon-Kutasov duality including the
${\cal N}=6$ case was proposed in \cite{Kapustin:2010mh}, which assumed
one conjecture to be proven. In particular, their conjecture gives a
formula for the phase differences of the dual pairs. We will explicitly
confirm their claim in our examples below.

%%%%%%%%%%%%%%%%%%%%%%%%%%%%%%%%%%%%%%
\subsubsection*{Seiberg duality for $\boldsymbol{N_1=1}$}

For $N_1=1$, the duality relation \eqref{Seibergduality} reads
\begin{align}
 U(1)_k\times U(N_2)_{-k} = U(1)_{-k}\times U(2+|k|-N_2)_k\ .
 \label{1N2Seibergduality}
\end{align}
In this case, we can actually prove that the integral representation
\eqref{ABJpart_integral} indeed gives identical results for the dual
pair, up to a phase.  Let us rewrite the $(1, N_2)$ partition function
given in \eqref{ABJN11} in the following form
\begin{align}
 Z_{\rm ABJ}(1, N_2)_k&=(2|k|)^{-1} Z_{\rm CS}^0(N_2-1)_k\, I(1,N_2)_k\, 
 e^{i\theta(1,N_2)_k}\ .\label{ABJN11(2)}
\end{align}
Here, $Z_{\rm CS}^0(M)_k$ is the Chern-Simons (CS) partition function
\begin{align}
 Z_{\rm CS}^0(M)_k=|k|^{-{M\over 2}}
 \prod_{j=1}^{M-1} \left(2\sin{\pi j\over |k|}\right)^{M-j}
 \label{CSMM0}
\end{align}
which is essentially the same as \eqref{CSMM} up to a phase due to
difference in the framing \cite{Marino:2011nm}.  Moreover,
\begin{align}
I(1,N_2)_k:&=-{1\over 2\pi i}
  \int_{I} {\pi \, ds\over \sin(\pi s)}
\prod_{l=1}^{N_2-1}\tan\left({(s+l)\pi\over |k|}\right)\ ,
\label{1N2integral}\\
  \theta(1,N_2)_k&: =-{\pi\over 6k}N_2(N_2-1)(N_2-2).
\end{align}
We can show that $Z_{\rm CS}^0(N_2-1)_k$ and $I(1,N_2)_k$ are separately
invariant under the Seiberg duality while the phase factor
$e^{i\theta(1,N_2)_k}$ gives a phase that precisely agrees with the one
given in \cite{Kapustin:2010mh}.

First, the invariance of $Z_{\rm CS}^0(N_2-1)_k$ is nothing but the
level-rank duality of the CS partition function, which means the
identity $Z_{\rm CS}^0(M)_k=Z_{\rm CS}^0(|k|-M)_k$.\footnote{A proof of
the level-rank duality can be found {\it e.g.}\ in Appendix B of
\cite{Kapustin:2010mh}.}  It is straightforward to see that this implies
that $Z_{\rm CS}^0(N_2-1)_k$ is invariant under the Seiberg duality
\eqref{1N2Seibergduality}.
Second, the phase difference between the
dual theories \eqref{1N2Seibergduality} is
\begin{align}
 \theta(1,N_2)_{k}-\theta(1,2+|k|-N_2)_{-k}
 =
 \pi \left[\kappa\left(-\frac{1}{6}k^2-\frac{1}{2}N_2^2+N_2-\frac{1}{3}\right)+\frac{1}{2}k(N_2-1)\right].
\end{align}
One can show that this phase difference is exactly the same as the one
given in \cite{Kapustin:2010mh}.

Now let us move on to the most nontrivial part, {\it i.e.}, the
invariance of the integral \eqref{1N2integral} under the Seiberg
duality.  One can show that, despite appearances, the integrand is
actually the same function for the dual theories
\eqref{1N2Seibergduality} up to a shift in $s$.  Therefore, the contour
integral gives the same answer for the duals, if the contour is chosen
appropriately.  As explained in section \ref{mainresults}, the integrand
has perturbative (P) poles coming from $\pi\over \sin(\pi s)$ and
non-perturbative (NP) poles coming from the product factor $\prod_l \tan$.  Although the
integrand remains the same under the Seiberg duality, the interpretation
of its poles gets interchanged; {\it i.e.}, a P pole in the original
theory is interpreted as a NP pole in the dual theory, and {\it vice
versa.}  We will see this explicitly in examples below, relegating the
general proof to Appendix \ref{Appendix_SD}\@.

The integrand of \eqref{1N2integral} is an antiperiodic (periodic)
function with $s\cong s+|k|$ for odd (even) $k$, and the P and NP poles
occur on the real $s$ axis in bunches with this periodicity.  The
prescription for the contour is to take it to go to the left of one of
such bunches. In Appendix \ref{Appendix_SD}, we show that this means
that
\begin{align}
 \eta=\begin{cases}
       0_+ & \qquad \text{if}\quad  {|k|\over 2}-N_2+1\ge 0,\\
       -{|k|\over 2}+N_2-1+0_+ & \qquad \text{if}\quad  {|k|\over 2}-N_2+1\le 0.\\
      \end{cases}
\label{etaPrescription}
\end{align}
This is required for the Seiberg duality to work, but it is also
necessary for the ABJ partition function to be analytic in $k$, which
is clearly the case for the original expression \eqref{ABJMM}.  In the
weak coupling regime $|g_s|=|2\pi i/k|\ll 1$, the NP poles are far away
from the origin (distance $\sim 1/|g_s|\sim |k|$) and we can safely take
$\eta=0_+$. However, as we decrease $|k|$ continuously, the NP poles
come closer to the origin and, eventually, at some even $|k|$, one of the
NP poles that was in the $s>0$ region reaches $s=0$.  As we further
decrease $|k|$ continuously, this NP pole enters the $s<0$ region. In
order for the partition function to be analytic in $k$, one needs to
increase the value of $\eta$ so that this NP pole does not move across
the contour $I$ but stays to the right of it.

%%%%%%%%%%%%%%%%%%%%%%%%%%%%%%%%%%%%%%
\subsubsection*{Odd $\boldsymbol{k}$ case}

\begin{figure}[h!]
\begin{center}
 \begin{tabular}{cc}
 \includegraphics[height=5cm]{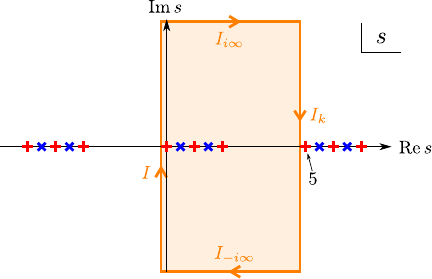} &
  \includegraphics[height=5cm]{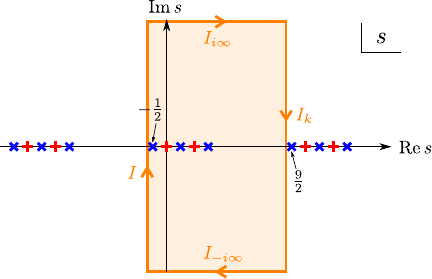} \\ 
  (a) $|k|=5,N_2=3$ &
  (b) $|k|=5,N_2=4$ \\[2ex]
  \includegraphics[height=5cm]{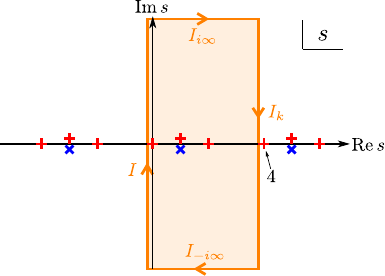} &
  \includegraphics[height=5cm]{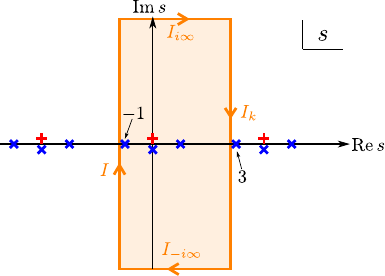}
\\
  (c) $|k|=4,N_2=2$  &
  (d) $|k|=4,N_2=4$
 \end{tabular}
 \caption{\sl The integration contour $C=I\cap I_{i\infty}\cap I_k\cap
 I_{-i\infty}$ (clockwise) and poles, for various values of $k,N_2$. (a)
 and (b) are Seiberg duals of each other and so are (c) and (d).  The
 contour $I_k$ is parallel to $I$ and shifted by $k$, and the contours
 $I_{i\infty}$ and $I_{-i\infty}$ are at infinity.  ``$+$'' (red)
 denotes the P pole and ``$\times$'' (blue) the NP pole.  Some poles and
 zeros are shown slightly above or below the real $s$ axis, but this is
 for the convenience of presentation and all poles and zeros are on the
 real $s$ axis.  The choices of the parameter $\eta$ for the contour $I$
 are $\eta=0_+$ for (a) and (c), $\eta=\half+0_+$ for (b), and
 $\eta=1+0_+$ for (d). \label{contour2}}
\end{center}
\end{figure}

The integral (\ref{1N2integral}) for odd $k$ is equal to the following
contour integral
\begin{align}
 I(1,N_2)_{k}=-{1\over 4\pi i}
  \int_{C} {\pi \, ds\over \sin(\pi s)}
 \prod_{l=1}^{N_2-1}\tan\left({(s+l)\pi\over |k|}\right)\ ,
\end{align}
where the integral contour $C$ is given by $C=I\cap I_{i\infty}\cap
I_k\cap I_{-i\infty}$ (clockwise), where the contour $I_k$ is parallel
to $I$ and shifted by $|k|$, and the contours $I_{i\infty}$ and
$I_{-i\infty}$ are at infinity; see Figure \ref{contour2}.  Note that
the anti-periodicity of the integrand allows us to write the line integral
(\ref{1N2integral}) as a closed contour integral, but the contour is
different from the tentative contour shown for the sake of sketchy illustration in Figures \ref{contour1} and \ref{NPpoles}.  By summing up pole residues inside $C$, one finds
\begin{align}
 I(1,N_2)_{k}
 &={1\over 2}
 \biggl[
 \sum_{n=0}^{|k|-N_2}(-1)^n\prod_{j=1}^{N_2-1}\tan{\pi(n+j)\over |k|}
 -|k|(-1)^{|k|-1\over 2}\sum_{n=1}^{N_2-1}
 (-1)^n\prod_{\substack{j=1\\ (\neq n)}}^{N_2-1}
 \tan{\pi({k\over 2}-n+j)\over |k|}
 \biggr].\label{mgzs23Nov12}
\end{align}
The first term comes from P poles and the second from  NP
poles. Although we prove the Seiberg duality in Appendix
\ref{Appendix_SD}, it is quite nontrivial that \eqref{mgzs23Nov12} gives
the same value for the dual pair \eqref{1N2Seibergduality}.

%\begin{figure}[h!]
%\centering
%\includegraphics[height=2.5in]{contourD.pdf}
%\hspace{.5cm}
%\includegraphics[height=2.5in]{contourE.pdf}
%\caption{\sl ``The integration contour'' $C=I\cap I_{\infty}\cap I_k\cap I_{-\infty}$ (clockwise) and poles. The left panel is for $k=5$ with $N_2=3$ and $4$, and the right panel is for $k=4+0_+$ with $N_2=2$ and $4$. The contour $I_k$ is parallel to $I$ and shifted by $k$, and the contours $I_{\infty}$ and $I_{-\infty}$ are at infinity. The choices of the parameter $\eta$ for the contour $I$ are $\eta=\half+0_+$ on the left panel and $\eta=1+0_+$ on the right panel. The detail of the pole structure is described in the main text.}
%\label{contour2}
%\end{figure}

Let us look at this in more detail in the following case:
\begin{align}
U(1)_5\times U(3)_{-5} = U(4)_5\times U(1)_{-5}\ .
\end{align}
Using the above formulas, we obtain the partition functions
of this dual pair which can be massaged into
\begin{align}
Z_{\rm ABJ}(1,3)_5&={1\over 50}\sin{\pi\over 5}\biggl[\underbrace{\tan{2\pi\over 5}\left(2\tan{\pi\over 5}+\tan{2\pi\over 5}\right)}_\text{P}\underbrace{-10\cot{\pi\over 5}}_\text{NP}\biggr]e^{-{\pi i\over 5}}\ ,\label{Z135}\\
Z_{\rm ABJ}(1,4)_{-5}&={1\over 50}\sin{\pi\over 5}\biggl[\underbrace{-10\cot{\pi\over 5}}_\text{P}\underbrace{+\tan{2\pi\over 5}\left(2\tan{\pi\over 5}+\tan{2\pi\over 5}\right)}_\text{NP}\biggr]e^{4\pi i\over 5}\ .\label{Z145}
\end{align}
These two indeed agree up to a phase and the phase difference agrees
with the conjecture made in \cite{Kapustin:2010mh}. Observe that the
contributions from the P and NP poles are interchanged under the
duality.  See Figure \ref{contour2}(a), (b) for the structure of the P
and NP poles in the two theories.
For discussion on the pole structure in more general cases, we refer the
reader to Appendix \ref{Appendix_SD}\@.

%%%%%%%%%%%%%%%%%%%%%%%%%%%%%%%%%%%%%%
\subsubsection*{Even $\boldsymbol{k}$ case}

The even $k$ case is technically a little more tricky.  Using a trick
similar to the one used in \cite{Okuyama:2011su}, the integral
(\ref{1N2integral}) for even $k$ can be shown to be equal to the following
contour integral
\begin{align}
 I(1,N_2)_{k}=-{1\over 2\pi i}
  \int_{C} {\pi \, ds\over \sin(\pi s)}\left(a-{s\over k}\right)\ 
 \prod_{l=1}^{N_2-1}\tan\left({(s+l)\pi\over k}\right),
\end{align}
where $a$ is an arbitrary constant.  For ${|k|\over
2}-N_2+1\ge 0$, we can evaluate this by summing over pole residues and obtain
\begin{align}
 &I(1,N_2)_{k}
 =\left(\sum_{n=0}^{{|k|\over 2}-N_2}+\sum_{n={|k|\over 2}}^{|k|-N_2}\right)
 \left(a-{n\over |k|}\right)(-1)^n\prod_{j=1}^{N_2-1}\tan{\pi(n+j)\over |k|}\notag\\
& \quad
 +\sum_{n=1}^{N_2-1}
 (-1)^{{|k|\over 2}-n}
 \left[-\left(a-{1\over 2}+{n\over |k|}\right)\sum_{\substack{j=1\\(j\neq n)}}^{N_2-1}
 {2\over \sin{2\pi \left({|k|\over 2}-n+j\right)\over |k|}}
 +{1\over \pi}
 \right]\prod_{\substack{j=1\\ (j\neq n)}}^{N_2-1}\tan{\pi(n+j)\over |k|}.\label{Ievenk}
\end{align}
The first line comes from P poles which are simple, while the second
line comes from double poles created by simple NP and P poles sitting on
top of each other. We note also that, despite its appearance, this expression does not depend on the constant $a$.
The expression of $I(1,N_2)_k$ for ${|k|\over 2}-N_2+1\le 0$ is more
lengthy and we do not present it, because the Seiberg duality proven in
Appendix \ref{Appendix_SD} guarantees that it can be obtained from
\eqref{Ievenk}.

Let us study in detail the following duality 
\be
U(1)_4\times U(2)_{-4} = U(4)_4\times U(1)_{-4}\ .
\ee
The partition functions of this dual pair yield
\begin{align}
Z_{\rm ABJ}(1,2)_4&={1\over 32}\biggl[\underbrace{1}_\text{P}\underbrace{-{2\over \pi}}_\text{P+NP}\biggr]\ ,\label{Z124}\\
Z_{\rm ABJ}(1,4)_{-4}&={1\over 32}\biggl[\underbrace{-{2\over \pi}}_\text{P+NP}\underbrace{+1}_\text{NP}\biggr]e^{\pi i}\ .\label{Z144}
\end{align}
These two agree up to a phase. The phase difference is again in
agreement with \cite{Kapustin:2010mh}.  The pole structure of the two
theories is shown in Figure \ref{contour2}(c), (d).  In the above,
``P+NP'' means the contribution from a double pole that comes from P and
NP poles on top of each other.  Again, the contributions from the P and
NP poles are interchanged under the duality.
Actually, in the even $k$ case, there is a subtlety in interpreting
simple poles as P or NP, but for details we refer the reader to Appendix
\ref{Appendix_SD}\@.

%%%%%%%%%%%%%%%%%%%%%%%%%%%%%%%%%%%%%%%%%%%%%%%%%%%%%%%%%%

\section{Conclusions and discussions}
\label{CandD}

In this paper, we have studied the partition function of the ABJ theory, {\it i.e.}, the ${\cal N}=6$ supersymmetric $U(N_1)_k\times U(N_2)_{-k}$ Chern-Simons-matter theory dual to M-theory on $AdS_4\times S^7/Z_k$ with a discrete torsion or type IIA string theory on $AdS_4\times CP^3$ with a NS-NS $B_2$-field turned on \cite{Aharony:2008gk}. More concretely, we have computed the ABJ partition function (\ref{ABJMM}) and found the expression (\ref{ABJpart_integral}) in terms of ${\rm min}(N_1,N_2)$-dimensional integrals as opposed to the original $(N_1+N_2)$-dimensional integrals. This generalizes the ``mirror description'' of the partition function of the ABJM theory \cite{Kapustin:2010xq} and may serve as the starting point for the ABJ generalization of the Fermi gas approach \cite{Marino:2011eh}.
We have taken an indirect approach: Instead of performing the eigenvalue integrals in (\ref{ABJMM}) directly, we have first calculated the partition function of the L(2,1) lens space matrix model (\ref{lensMM}) \emph{exactly} and found the expression (\ref{lensAns2}) as a product of $q$-deformed Barnes $G$-function and a generalization of multiple $q$-hypergeometric function. We have then performed the analytic continuation $N_2\to -N_2$ of the lens space partition function to obtain the ABJ partition function.
As checks we have shown that our main result (\ref{ABJpart_integral}) correctly reproduces perturbative expansions and in the $N_1=1$ case, {\it i.e.}, for the $U(1)_k\times U(N_2)_{-k}$ theories, the Seiberg duality indeed holds. In particular, we have uncovered that the perturbative and nonperturbative contributions to the partition function are interchanged under the Seiberg duality and derived, in the $N_1=1$ case, the formula for the phase difference of dual-pair partition functions conjectured in \cite{Kapustin:2010mh}. It is also worth remarking that the ABJ partition function (\ref{ABJpart_integral}) vanishes for $|N_1-N_2|>k$ in line with the conjectured supersymmetry breaking \cite{Bergman:1999na}.

As commented before, we note, however, that the analytic continuation is ambiguous and we have adopted a particular prescription that required {\it a posteriori\/} justification. Especially, our prescription involves an intermediate step, namely an infinite sum expression, (\ref{ABJpart_formalsum}) which is non-convergent and becomes singular for an even integer $k$. Although the integral representation (\ref{ABJpart_integral}) provides a regularization and an analytic continuation of the formal series (\ref{ABJpart_formalsum}) in the complex $q$-plane, it would be better if we could render every step of the calculation process well-defined. In this connection, it is somewhat dissatisfying that the $q$-hypergeometric structure enjoyed by the lens space partition function becomes obscured after the analytic continuation to the ABJ partition function. It might be that there is a better way to perform the analytic continuation that manifestly respects the $q$-hypergeometric structure and directly yields a finite sum expression for an integer $k$ without passing to the integral representation.

Although the successful test of the Seiberg duality for the $U(1)_k\times U(N_2)_{-k}$ theories provides  compelling evidence for our prescription, a general proof is clearly desired. In this regard, we note, as discussed in Section \ref{SD}, that the Seiberg duality acts on the $U(|N_1-N_2|)_k$ CS factor and the integral part separately. Namely, apart from a phase factor, the CS and the integral parts are respectively invariant under the duality, where the invariance of the former follows from the level-rank duality. Thus the general proof amounts to showing the invariance of the integral part, {\it i.e.}, the second line of (\ref{ABJpart_integral}). We leave this proof for a future work.

Following this work, there are a few more immediate directions to
pursue: It is straightforward to generalize our computation of the
partition function to Wilson loops \cite{Drukker:2008zx, Chen:2008bp,
Rey:2008bh, Drukker:2009hy, Cardinali:2012ru}. Indeed, we can proceed
almost in parallel with the case of the partition function for the most
part including the analytic continuation, although the computation
becomes inevitably more involved.  We hope to report on our progress in
this direction in the near future \cite{WIP2}.  It may also be possible
to apply our method to more general CSM theories with fewer
supersymmetries, provided that a similar analytic continuation works.
Meanwhile, we have stressed in the introduction that this work may have
significance to the study of higher spin theories, especially, in
connection to the recent ABJ triality conjecture \cite{Chang:2012kt}. As
mentioned towards the end of Section \ref{Examples}, it is in fact
feasible to analyze the $U(1)_k\times U(N_2)_{-k}$ and $U(2)_k\times
U(N_2)_{-k}$ partition functions at large $N_2$ and $k$
\cite{WIP1}. This may shed lights on the understanding of the ${\cal N}=6$
parity-violating Vasiliev theory on $AdS_4$.  In particular, for the
$U(1)_k\times U(N_2)_{-k}$ theory, the fact that the Seiberg duality
separately acts on the $U(N_2-1)$ CS and the integral
parts seems to suggest that it is only the integral part that may be dual to
the vector-like Vasiliev theory.

Last but not least, it is most important to gain, if possible, new physical and mathematical insights into the microscopic description of M-theory through all these studies. 
Although the ABJ(M) theory is a very useful and practical description of maximally supersymmetric 3d conformal field theories, the construction by Bagger-Lambert and Gustavsson based on a 3-algebra \cite{Bagger:2006sk, Gustavsson:2007vu} is arguably more insightful, suggesting potentially a new mathematical structure behind quantum membrane theory. What we envisage in this line of study is to search for a way to reorganize the ABJ(M) partition function in terms of the degrees of freedom that might provide an intuitive understanding of the $N^{3/2}$ scaling and suggest hidden structures behind the microscopic description of M-theory such as 3-algebras.

%%%%%%%%%%%%%%%%%%%%%%%%%%%%%%%%%%%%%%%%%%%%%%%%%%%%%%%%%%

%%%%%%%%%%%%%%%%%%%%%%%%%%%%%%%%%%%%%%%%%%%%%%%%%%%%%%%%%%

\section*{Acknowledgments}

We would like to thank Oren Bergman, Hiroyuki Fuji, Yoichi Kazama,
Sanefumi Moriyama, Vasilis Niarchos, Keita Nii, Kazutoshi Ohta, Shuichi
Yokoyama, Xi Yin, and Tamiaki Yoneya for comments and discussions.  The
work of HA was supported in part by Grant-in-Aid for Scientific Research
(C) 24540210 from the Japan Society for the Promotion of Science
(JSPS)\@.  The work of SH was supported in part by the Grant-in-Aid for
Nagoya University Global COE Program (G07).  The work of MS was
supported in part by Grant-in-Aid for Young Scientists (B) 24740159 from
the Japan Society for the Promotion of Science (JSPS)\@.

\appendix
%\newpage

\section{$\boldsymbol{q}$-analogs}
\label{Appendix_q-Analogs}

Roughly, a $q$-analog is a generalization of a quantity to include a new
parameter $q$, such that it reduces to the original version in the $q\to
1$ limit.  In this appendix, we will summarize definitions of various
$q$-analogs and their properties relevant for the main text.

\paragraph{$\boldsymbol{q}$-number:}  
For $z\in\bbC$, the $q$-number of $z$ is defined by
\begin{align}
 [z]_q:={1-q^z\over 1-q},\qquad 
\end{align}

\paragraph{$\boldsymbol{q}$-Pochhammer symbol:}

For $a\in\bbC$, $n\in\bbZ_{\ge 0}$, the $q$-Pochhammer symbol $(a;q)$ is defined by
\begin{align}
 (a;q)_n&:= \prod_{k=0}^{n-1}(1-aq^k)
 =(1-a)(1-aq)\cdots(1-aq^{n-1})
 ={(a;q)_\infty\over (aq^n;q)_\infty}.
\end{align}
For $z\in\bbC$, $(a;q)_z$ is defined by the last expression:
\begin{align}
 (a;q)_z&:=
 {(a;q)_\infty\over (aq^z;q)_\infty}
 =\prod_{k=0}^\infty{1-aq^k\over 1-aq^{z+k}}
 .\label{maux20Sep12}
\end{align}
This in particular means
\begin{align}
 (a;q)_{-z}&={1\over (aq^{-z};q)_z}.
\end{align}
For $n\in \bbZ_{\ge 0}$,
\begin{align}
 (a;q)_{-n}&={1\over (aq^{-n};q)_n}={1\over \prod_{k=1}^n(1-a/q^k)}.
\end{align}

Note that the $q\to 1$ limit of the $q$-Pochhammer symbol is not the usual
Pochhammer symbol but only up to factors of $(1-q)$:
\begin{align}
  \lim_{q\to 1}{(q^a;q)_n\over (1-q)^n}&=
 a(a+1)\dots(a+n-1).
\end{align}

We often omit the base $q$ and simply write $(a;q)_\nu$ as
$(a)_\nu$.\footnote{We will not use the symbol $(a)_\nu$ to denote the
usual Pochhammer symbol.}

Some useful relations involving $q$-Pochhammer symbols are
\begin{align}
 (a)_\nu&={(a)_z\over (aq^\nu)_{z-\nu}}=(a)_z (aq^z)_{\nu-z},\label{hcey4Sep12}\\
 (q)_\nu&=(1-q)^{\nu}\Gamma_q(\nu+1),\label{kjzv20Sep12}\\
 (q^\mu)_\nu&={(q)_{\mu+\nu-1}\over (q)_{\mu-1}} =(1-q)^\nu{\Gamma_q(\mu+\nu)\over \Gamma_q(\mu)},\\
 (aq^\mu)_\nu&
 =(aq^\mu)_{z-\mu}(aq^z)_{\mu+\nu-z}
 ={(aq^\mu)_z\over (aq^{\mu+\nu})_{z-\nu}}
 ={(aq^z)_{\mu+\nu-z}\over (aq^z)_{\mu-z}},
 \label{jwdu20Sep12}
\end{align}
where $\mu,\nu,z\in\bbC$ and $\Gamma_q(z)$ is the $q$-Gamma function
defined below. For $n\in\bbZ$, we have the following formulae which
``reverse'' the order of the product in the $q$-Pochhammer symbol:
\begin{align}
 (aq^z)_n&=(-a)^n q^{zn+\half n(n-1)}(a^{-1}q^{1-n-z})_n,\label{jzmp20Sep12}\\
 (\pm q^{-n})_n&=(\mp 1)^n q^{-\half n(n+1)}(\pm q)_n.\label{hcit4Sep12}
\end{align}
If $\nu=n+\epsilon$ with $|\epsilon|\ll 1$, the correction to this is of
order $\cO(\epsilon)$:
\begin{align}
 (aq^z)_{n+\epsilon}&=(-a)^nq^{zn+{1\over 2}n(n-1)}(a^{-1}q^{1-n-z})_n(1+\cO(\epsilon)),\qquad a\neq 1.
\end{align}
Here we assumed that $a\neq 1$ and $a-1\gg \cO(\epsilon)$,
%If we furthermore have $a=1$ and $z=m+\delta$ ($\delta\ll 1$), there can
%be subtlety coming from $1-q^0$ and we can have large correction that depends
%on the ratio $\epsilon\over\delta$. (see notes on 9/20/2012).

\paragraph{$\boldsymbol{q}$-factorials:} For $n\in\bbZ_{\ge 0}$, the $q$-factorial is given by
\begin{align}
 [n]_q!:= [1]_q[2]_q\cdots[n]_q={(q)_n\over (1-q)^n},\qquad [0]_q!=1,
\qquad
 [n+1]_q!=[n]_q[n-1]_q!~.
\end{align}
\paragraph{$\boldsymbol{q}$-Gamma function:} For $z\in\bbC$, the $q$-Gamma function $\Gamma_q(z)$ is defined by
\begin{align}
 \Gamma_q(z+1)&:= (1-q)^{-z}\prod_{k=1}^\infty {1-q^{k}\over
 1-q^{z+k}}.
\end{align}
The $q$-Gamma function satisfies the following relations:
\begin{align}
 \Gamma_q(z)&
 =(1-q)^{1-z}{(q)_\infty\over (q^z)_\infty}
 =(1-q)^{1-z}(q)_{z-1},\\
 \Gamma_q(z+1)&=[z]_q\Gamma_q(z),\\
 \Gamma_q(1)&=\Gamma_q(2)=1,\qquad
 \Gamma_q(n)=[n-1]_q! \quad (n\ge 1).
\end{align}
The behavior of $\Gamma_{q}(z)$ near non-positive integers is
\begin{align}
 \Gamma_q(-n+\epsilon)
 &={(-1)^{n+1}(1-q)q^{\half n(n+1)}\over \Gamma_q(n+1)\,\log q}
 {1\over \epsilon}+\cdots,\qquad
  \Gamma_q(n+1)=[n]_q!~,\quad %\checkmark
% &={(-1)^{n+1}(1-q)^{n+1}q^{\half n(n+1)}\over (q)_n\log q}
% {1\over \epsilon}+\cdots\\
\end{align}
where $n\in\bbZ_{\ge 0}$, and $\epsilon\to 0$.   As $q\to 1$, this reduces to the formula for the
ordinary $\Gamma(z)$,
\begin{align}
 \Gamma(-n+\epsilon)&={(-1)^{n}\over \Gamma(n+1)}{1\over \epsilon}+\cdots,\qquad
 \Gamma(n+1)=n!~.
\end{align}

\paragraph{$\boldsymbol{q}$-Barnes $\boldsymbol{G}$ function:}
For $z\in\bbC$, the $q$-Barnes $G$ function is defined by \cite{Nishizawa}
\begin{align}
 G_2(z+1;q)&:=(1-q)^{-{1\over 2}z(z-1)}
 \prod_{k=1}^\infty 
 \biggl[\biggl({1-q^{z+k}\over 1-q^k}\biggr)^k(1-q^k)^z
 \biggr].
\end{align}
Some of its properties are
\begin{gather}
 G_2(1;q)=1,\qquad G_2(z+1;q)=\Gamma_q(z)G_2(z),\\
 G_2(n;q)
 =\prod_{k=1}^{n-1}\Gamma_q(k)
 =\prod_{k=1}^{n-2} [k]_q!
 =(1-q)^{-\half(n-1)(n-2)}\prod_{j=1}^{n-2}(q)_j
 =\prod_{k=1}^{n-2}[k]_q^{n-k-1},\\
 \prod_{1\le A<B\le n}(q^A-q^B)=
 q^{{1\over 6}n(n^2-1)}(1-q)^{\half n(n-1)}G_2(n+1;q).
 \label{ehwt6Nov12}
\end{gather}
The behavior of $G_2(z;q)$ near non-positive integers is
\begin{align}
 G_2(-n+\epsilon;q)
 ={(-1)^{\half (n+1)(n+2)}G_2(n+2;q)\,(\log q)^{n+1}
 \over q^{{1\over 6}n(n+1)(n+2)} (1-q)^{n+1}}\epsilon^{n+1}+\cdots,
% \quad G_2(n+2;q)=\prod_{k=1}^n[k]_q!,
 \label{ih4Nov12}
\end{align}
where $n\in \bbZ_{\ge 0}$, and $\epsilon\to 0$. As $q\to 1$, this
reduces to the formula for the ordinary $G_2(z)$,
\begin{align}
 G_2(-n+\epsilon)
 &=(-1)^{\half n(n+1)}G_2(n+2)\epsilon^{n+1}+\cdots.
% \quad G_2(n+2)=\prod_{k=1}^n k!~
 \label{il4Nov12}
\end{align}

\paragraph{$\boldsymbol{q}$-digamma and $\boldsymbol{q}$-polygamma functions}

The $q$-digamma function $\psi_q(z)$ and $q$-polygamma function
$\psi^{(n)}_q(z)$, $n\in\bbZ_{\ge 0}$, are defined by
\begin{align}
 \psi_q(z):=\partial_z \ln \Gamma_q(z),\qquad
 \psi^{(n)}_q(z):=\partial_z^n \psi_q(z)=
 \partial_z^{n+1}\ln \Gamma_q(z).
\end{align}
From the definition of $\Gamma_q(z)$, it straightforwardly follows that
\begin{align}
 \psi_q(z)&=-\log(1-q)+\sum_{n=0}^\infty {q^{n+z}\over 1-q^{n+z}}\ln q,\qquad
 \psi^{(1)}_q(z)=\sum_{n=0}^\infty {q^{n+z}\over (1-q^{n+z})^2}\ln^2 q.
\end{align}

\paragraph{$\boldsymbol{q}$-hypergeometric function (basic hypergeometric series):}

The $q$-hypergeometric function, or the basic hypergeometric series
with base $q$, is defined by \cite{Gasper-Rahman}
\begin{align}
 _{r}\phi_s\left({a_1,\dots,a_r\atop b_1\dots,b_s};q,z\right)
 :=\sum_{n=0}^\infty 
 {(a_1)_n\cdots (a_r)_n\over (q)_n (b_1)_n\cdots (b_s)_n}
 \left[(-1)^nq^{n\choose 2}\right]^{1+s-r}z^n.
\end{align}
In particular,
for
$r=k+1,s=k$,
\begin{align}
 _{k+1}\phi_k\left({a_1,\dots,a_{k+1}\atop b_1\dots,b_k};q,z\right)
 =\sum_{n=0}^\infty {(a_1)_n\cdots (a_{k+1})_n\over (b_1)_n\cdots (b_k)_n}
 {z^n\over (q)_n}.
\end{align}

\section{Lens space matrix model}
\label{Appendix_LensMM}

The partition function for the lens space matrix model was defined in
\eqref{lensMM}.  Here, we explicitly carry out the integral and write
the result in a simple closed form as given in \eqref{lensAns1},
\eqref{lensAns2}.  The following computation can be thought of as a
generalization of the matrix integration technique using Weyl's
denominator formula (see for example \cite{Aganagic:2002wv,
Marino:2011nm}), explicitly worked out.

First, we note that the 1-loop determinant part can be 
reduced to a single Vandermonde determinant
by shifting the integration variables as $\mu_j\to
\mu_j-{i\pi\over 2},\nu_a\to\nu_a+{i\pi\over 2}$, as follows:
\begin{align}
\begin{split}
 &\Delta_{\rm sh}(\mu)\Delta_{\rm sh}(\nu)\Delta_{\rm ch}(\mu,\nu)\\
 &=
 \prod_{j<k} e^{-{\mu_j+\mu_k\over 2}}(e^{\mu_j}-e^{\mu_k})
 \prod_{a<b} e^{-{\nu_a+\nu_b\over 2}}(e^{\nu_a}-e^{\nu_b})
 \prod_{j,a} e^{-{\mu_j+\nu_k\over 2}}(e^{\mu_j}+e^{\nu_a})\\
 &\to
 \prod_{j<k} e^{-{\mu_j+\mu_k\over 2}}(e^{\mu_j}-e^{\mu_k})
 \prod_{a<b} e^{-{\nu_a+\nu_b\over 2}}(e^{\nu_a}-e^{\nu_b})
 \prod_{j,a}
e^{-{i\pi\over 2}}e^{-{\mu_j+\nu_k\over 2}}(e^{\mu_j}-e^{\nu_a})
 \\
 &=
 e^{-{i\pi\over 2}N_1N_2-{N-1\over 2}(\sum_j \mu_j+\sum_a \nu_a)}\Delta(\mu,\nu),
\end{split}
\end{align}
where $N:= N_1+N_2$ and $\Delta(\mu,\nu)$ is the Vandermonde
determinant  for $(\mu_j,\nu_a)$ which can be evaluated as
\begin{align}
\Delta(\mu,\nu) 
 &:=
 \prod_{j<k} (e^{\mu_j}-e^{\mu_k})
 \prod_{a<b} (e^{\nu_a}-e^{\nu_b})
 \prod_{j,a} (e^{\mu_j}-e^{\nu_a})\notag\\
 &=\sum_{\sigma\in S_{N}} (-1)^\sigma 
 e^{ \sum_{j=1}^{N_1} (\sigma(j)-1)\mu_j
 +\sum_{a=1}^{N_2}(\sigma(N_1+a)-1)\nu_a }.
 \label{itho1Nov12}
\end{align}
Here, $S_N$ is the permutation group of length $N$ and $(-1)^\sigma$ is
the signature of $\sigma\in S_N$. Because each term in
\eqref{itho1Nov12} is an exponential whose exponent is linear in
$\mu_j,\nu_a$, the integral in \eqref{lensMM} is trivial
Gaussian. After carrying out the $\mu_i,\nu_a$ integrals and massaging
the result a little bit, we obtain
\begin{align}
\begin{split}
  Z_{\text{lens}}(N_1,N_2)_k&
 =\cN_{\text{lens}}(-1)^{{1\over 2}N_1(N_1+1)+{1\over 2}N_2(N_2+1)+N_1N_2}
 e^{-{g_s\over 6}N(N+1)(N+2)}
 \left({g_s\over 2\pi}\right)^{N\over 2}
 Z_{\text{lens}}^0,\\
 Z_{\text{lens}}^0&:=
 \sum_{\sigma,\tau\in S_{N}}(-1)^{\sigma+\tau}
 e^{g_s\sum_{A=1}^N\sigma(A)\tau(A)
 +{i\pi\over 2}\left(\sum_{A=1}^{N_1}-\sum_{A=N_1+1}^{N_1+N_2}\right)(\sigma(A)+\tau(A))}.
\end{split} 
\label{lrdr25Aug12}
\end{align}
Note that the summation over $\tau$ in \eqref{lrdr25Aug12} can be
written in terms of a determinant as
\begin{align}
 Z_{\text{lens}}^0(N_1,N_2)_k&
 =\sum_\sigma (-1)^\sigma 
 e^{
 {i\pi\over 2}\left(\sum_{j=1}^{N_1}-\sum_{j=N_1+1}^{N_1+N_2}\right)\sigma(j)}\det W(\sigma),\label{lreb25Aug12}\\
 W(\sigma)_{AB}&:=\begin{cases}
		 e^{(g_s\sigma(A)+{i\pi\over 2})B}& \qquad (1\le A\le N_1),\\
		 e^{(g_s\sigma(A)-{i\pi\over 2})B}& \qquad (N_1+1\le A\le N).
		\end{cases}
\end{align}
The matrix $W$ is essentially a Vandermonde matrix and its determinant can be
evaluated using the formula
\begin{align}
 \det[(x_A)^B]=\biggl(\prod_{A=1}^N x_A\biggr)^N \prod_{1\le A<B\le N}
 (x_A^{-1}-x_B^{-1})
\end{align}
as follows:
\begin{align}
 \det W(\sigma)
 &=e^{Ng_s\sum_{A=1}^N \sigma(A)+{i\pi\over 2}N(N_1-N_2)}
 \prod_{j<k}(e^{-g_s\sigma(j)-{i\pi\over 2}}-e^{-g_s\sigma(k)-{i\pi\over 2}})\notag\\
 &\qquad\qquad\times
 \prod_{a<b}(e^{-g_s\sigma(a)+{i\pi\over 2}}-e^{-g_s\sigma(b)+{i\pi\over 2}})
 \prod_{j,a}(e^{-g_s\sigma(j)-{i\pi\over 2}}-e^{-g_s\sigma(a)+{i\pi\over 2}})
 \notag\\
 &=e^{{i\pi\over 4}(N_1(N_1+1)-N_2(N_2+1)-2N_1N_2)}
 e^{{g_s\over 2} N^2(N+1)}
 \prod_{j<k}(e^{-g_s\sigma(j)}-e^{-g_s\sigma(k)})
 \notag\\
 &\qquad\qquad\times
 \prod_{a<b}(e^{-g_s\sigma(a)}-e^{-g_s\sigma(b)})
 \prod_{j,a}(e^{-g_s\sigma(j)}+e^{-g_s\sigma(a)}).
\end{align}
Plugging this into \eqref{lrdr25Aug12} and \eqref{lreb25Aug12}, the
expression for $Z_{\text{lens}}$ is
\begin{align}
 Z_{\text{lens}}(N_1,N_2)_k
 &=\cN_{\text{lens}}\left({g_s\over 2\pi}\right)^{N\over 2}
 (-1)^{\half N_1(N_1+1)}
 q^{-{1\over 3}N(N^2-1)}
 \sum_{\sigma\in S_N} (-1)^{\sigma+
 \sum_{A=1}^{N_1}\sigma(A)}\notag\\
 &\qquad 
 \times\prod_{j<k}(q^{\sigma(j)}-q^{\sigma(k)})
 \prod_{a<b}(q^{\sigma(a)}-q^{\sigma(b)})
 \prod_{j,a}(q^{\sigma(j)}+q^{\sigma(a)})\label{jims29Aug12}
\end{align}
where $q=e^{-g_s}$.

We can rewrite \eqref{jims29Aug12} in a simpler form as follows.
$\sigma$ is a permutation of length $N=N_1+N_2$.  Let us take its first
$N_1$ entries $\sigma(1),\sigma(2),\dots, \sigma(N_1)$, permute them
into increasing order, and call them $C_1,\dots,C_{N_1}$
($C_1<\cdots<C_{N_1}$).  Similarly, we take the last $N_2$ entries
$\sigma(N_1),\dots, \sigma(N)$, permute them into increasing order, and
call them $D_1,\dots,D_{N_2}$ ($D_1<\cdots<D_{N_2}$). Let the signature
for the permutation to take $(C_1,\dots,C_{N_1})$ to
$(\sigma(1),\dots,\sigma(N_1))$ be $(-1)^C$ and the signature for the
permutation to take $(D_1,\dots,D_{N_2})$ to
$(\sigma(N_1+1),\dots,\sigma(N))$ be $(-1)^D$.  Namely,
\begin{align}
 (-1)^C:= \sign\begin{pmatrix}
	   C_1&\dots &C_{N_1}\\
	   \sigma(1)&\dots &\sigma(N_1)
	  \end{pmatrix},\qquad
 (-1)^D:=\sign\begin{pmatrix}
	   D_1&\dots &D_{N_2}\\
	   \sigma(N_1+1)&\dots &\sigma(N)
	  \end{pmatrix}.\label{kznq29Aug12}
\end{align}
Then the factors in
\eqref{jims29Aug12} can be rewritten as
\begin{align}
  \prod_{j<k}(q^{\sigma(j)}-q^{\sigma(k)}) =(-1)^C \prod_{C_j<C_k}(q^{C_j}-q^{C_k})
 ,\quad
  \prod_{a<b}(q^{\sigma(a)}-q^{\sigma(b)}) =(-1)^D \prod_{D_a<D_b}(q^{D_a}-q^{D_b})\label{ldve29Aug12}.
\end{align}
These relations are easy to see by looking at the left hand side as
Vandermonde determinants.  Also, note that
\begin{align}
 \sign\begin{pmatrix}
       1&\dots&N_1&N_1+1&\dots&N\\
       C_1&\dots&C_{N_1}&D_1&\dots&D_{N_2}
      \end{pmatrix}
 =(-1)^{\half N_1(N_1+1)+\sum_{A=1}^{N_1}\sigma(A)}.\label{kzkp29Aug12}
\end{align}
This is seen as follows. First, let us permute $(C_1,\dots, C_{N_1})$
into $(C_{N_1},\dots,C_1)$, which gives $(-1)^{\half N_1(N_1-1)}$.  Next,
let us permute $(C_{N_1},\dots, C_1,D_1,\dots, D_{N_2})$ into
$(1,\dots,N)$, starting by moving $C_{N_1}$ to the correct position.
For this, $C_{N_1}$ commute through other $C_{N_1}-1$ numbers to its
right, giving $(-1)^{C_{N_1}-1}$.  Next, we move $C_{N_1-1}$ to the
correct position, which gives $(-1)^{C_{N_1-1}-1}$. We keep doing this
until we get $(1,\dots,N)$.  In the end, we obtain
$(-1)^{\sum_{j=1}^{N_1} (C_j-1)}=(-1)^{\sum_{A=1}^{N_1}
\sigma(A)-N_1}=(-1)^{\sum_{A=1}^{N_1} \sigma(A)+N_1}$.  Combining this
with the previous factor, we obtain \eqref{kzkp29Aug12}.  Eqs.\
\eqref{kznq29Aug12} and \eqref{kzkp29Aug12} mean that
\begin{align}
 (-1)^\sigma=(-1)^{C+D+\half N_1(N_1+1)+\sum_{A=1}^{N_1}\sigma(A)}.\label{louk29Aug12}
\end{align}
Plugging \eqref{ldve29Aug12} into \eqref{jims29Aug12} and using
\eqref{louk29Aug12}, we obtain the following nice concise formula for
the partition function for the lens space matrix model:
\begin{align}
 Z_{\text{lens}}(N_1,N_2)_k
 &=
 i^{-{\kappa\over 2}(N_1^2+N_2^2)}
 \left({g_s\over 2\pi}\right)^{N\over 2}
 q^{-{1\over 3}N(N^2-1)}
 \notag\\
 &\qquad
 \times \sum_{(\cN_1,\,\cN_2)}
 \prod_{C_j<C_k} (q^{C_j}-q^{C_k})
 \prod_{D_a<D_b} (q^{D_a}-q^{D_b})
 \prod_{C_j,D_a} (q^{C_j}+q^{D_a}),\label{ihsh28Aug12}
\end{align}
which is the expression presented in \eqref{lensAns1}.  Here,
$\sum_{(\cN_1,\,\cN_2)}$ means summation over different ways to
decompose $\{1,2,\dots,N_1+N_2\}$ into two disjoint sets $\cN_1$ and
$\cN_2$ with $\#\cN_1=N_1$, $\#\cN_2=N_2$. Their elements are
\begin{align}
 \cN_1&=\{C_1,C_2,\dots,C_{N_1}\},\qquad C_1<C_2<\dots<C_{N_1},\\
 \cN_2&=\{D_1,D_2,\dots,D_{N_2}\},\qquad D_1<D_2<\dots<D_{N_2}.
\end{align}

Note that, using the identity \eqref{ehwt6Nov12}, Eqn.\
\eqref{ihsh28Aug12} can also be rewritten as
\begin{align}
 Z_{\text{lens}}(N_1,N_2)_k
 &=
 i^{-{\kappa\over 2}(N_1^2+N_2^2)}
 \left({g_s\over 2\pi}\right)^{N\over 2}
 q^{-{1\over 6}N(N^2-1)}(1-q)^{\half N(N-1)}G_2(N+1;q)\,S(N_1,N_2),
 \label{gpmz3Sep12}
 \\
 S(N_1,N_2)&= \sum_{(\cN_1,\,\cN_2)}
 \prod\limits_{C_j<D_a}{q^{C_j}+q^{D_a}\over q^{C_j}-q^{D_a}}
 \prod\limits_{D_a<C_j}{q^{D_a}+q^{C_j}\over q^{D_a}-q^{C_j}},
 \label{msmi31Aug12}
\end{align}
which is the expression presented in \eqref{lensAns2}.

\section{Analytic continuation to ABJ matrix model}
\label{Appendix_AnalCont}

Here, we will obtain the ABJ matrix model partition function by
analytically continuing the lens space matrix model partition
function  \eqref{gpmz3Sep12} under $N_2\to -N_2$.

\subsection{Normalization}
\label{Appendix_Norm}

It has been shown \cite{Marino:2009jd} that the partition functions for
the lens space and ABJ theories agree order by order in perturbation
theory upon analytic continuing in the rank as $N_2\to -N_2$.  Our
strategy is to apply this analytic continuation to the lens space
partition function to obtain the exact expression for the ABJ partition
function.  However, in order to analytically continue the partition
functions, not just their perturbative expansion, we must properly
normalize them, which is what we discuss first.

Because we already know \cite{Marino:2009jd} that the analytic
continuation works perturbatively, all we have to do is to match the
tree level part of the partition function.  In the weak coupling limit
$g_s\to 0$, the lens space partition function \eqref{lensMM}
reduces to
\begin{align}
 Z_{\text{lens,tree}}
 &:=
 Z_{\text{lens}}(g_s\to 0)\notag\\
 &=
 2^{2N_1N_2}\cN_{\text{lens}}
 \int \prod_{j} {d\mu_j\over 2\pi}
 \prod_{a}{d\nu_a\over 2\pi}
 \prod_{j<k}(\mu_j-\mu_k)^2
 \prod_{a<b}(\nu_a-\nu_b)^2
 e^{-{1\over 2g_s}(\sum_j \mu_j^2+\sum_a \nu_a^2)}.\label{fyi3Nov12}
\end{align}
This is essentially the product of two copies of Gaussian matrix model
partition function:
\begin{align}
 Z_{\text{lens,tree}}=
 i^{-{\kappa\over 2}(N_1^2+N_2^2)} {2^{2N_1N_2}\over N_1! N_2!}
 V(N_1,g_s)V(N_2,g_s),\label{niev2Nov12}
\end{align}
where $V(n,g_s)$ is the $U(n)$ Gaussian matrix model integral,
\begin{align}
 V(n,g_s):=
 \int \prod_{j=1}^{n}{d\lambda_j\over 2\pi}
 \Delta(\lambda)^2
 e^{-{1\over 2g_s}\sum_{j=1}^n \lambda_j^2},\qquad
 \Delta(\lambda)=\prod_{1\le j<k\le n}(\lambda_j-\lambda_k).\label{jrqv3Sep12}
\end{align}
$V(n,g_s)$ can be computed explicitly as
\cite{Mehta}
\begin{align}
 V(n,g_s)
 = g_s^{n^2\over 2}(2\pi)^{-{n\over 2}}G_2(n+2),
 \label{juji2Nov12}
\end{align}
where $G_2(z)$ is the (ordinary) Barnes function.
In the present case we have $g_s={2\pi i\over k}$ and the integral
\eqref{jrqv3Sep12} is the Fresnel integral.  Similarly, the ABJ partition
function \eqref{ABJMM} reduces in the weak coupling limit to
\begin{align}
 Z_{\text{ABJ,tree}}
 \equiv
 Z_{\text{ABJ}}(g_s\to 0)
 =
 i^{-{\kappa\over 2}(N_1^2-N_2^2)} {2^{-2N_1N_2}\over N_1! N_2!} 
 V(N_1,g_s)V(N_2,-g_s).\label{gxyb3Nov12}
\end{align}
Note that
\begin{align}
 &V(n,-g_s)
 =
 (-g_s)^{n^2\over 2}(2\pi)^{-{n\over 2}}G_2(n+2)
 = i^{-\kappa n^2} g_s^{n^2\over 2}(2\pi)^{-{n\over 2}}G_2(n+2)
 = i^{-\kappa n^2}V(n,g_s).
 \label{gyjn3Nov12}
\end{align}
In the second equality, we used the fact that, because $g_s=2\pi i/k$,
the Gauss integrals we are doing are actually Fresnel integrals and
therefore
\begin{align}
 (\pm g_s)^{\half}=\sqrt{2\pi\over |k|}\,i^{\pm {\kappa\over 2}}.
 \label{gzzx3Nov12}
\end{align}
Using \eqref{gyjn3Nov12}, the tree level ABJ partition function
\eqref{gxyb3Nov12} can be written as
\begin{align}
 Z_{\text{ABJ,tree}}
 =
 i^{-{\kappa\over 2}(N_1^2+N_2^2)} {2^{-2N_1N_2}\over N_1! N_2!} 
 V(N_1,g_s)V(N_2,g_s).\label{fhsv3Nov12}
\end{align}
Looking at \eqref{niev2Nov12} and \eqref{fhsv3Nov12}, one may think that
$Z_{\text{lens}}$ is analytically continued to $Z_{\text{ABJ}}$ under
$N_2\to -N_2$.  However, this does not work because $N_2!=\Gamma(N_2+1)$
and $V(N_2,g_s)$ do not transform in the right way under $N_2\to -N_2$.

To find the correct way to normalize partition function, we observe that
the Gaussian matrix model \eqref{jrqv3Sep12} can be thought of as coming
from gauge fixing the ``ungauged'' Gaussian matrix integral,
\begin{align}
 \Vh(n,g_s)
 :=
 \int d^{n^2}\!M\, e^{-{1\over 2g_s}\tr M^2}
% =\int d^{n^2}\!M\, e^{{ik\over 4\pi}\tr M^2}
% =i^{n^2\over 2}\left({4\pi^2\over k}\right)^{n^2\over 2}
 =(2\pi g_s)^{n^2\over 2}
 ,
 \label{lbym22Sep12}
\end{align}
to the eigenvalue basis.  Our claim is that it is such ungauged matrix
integrals that should be used for analytic continuation between lens
space and ABJ theories.  Let us make this statement more precise.  Note
that the relation between the ungauged Gaussian matrix integral
\eqref{lbym22Sep12} and its gauge-fixed version \eqref{jrqv3Sep12},
\eqref{juji2Nov12} is
\begin{align}
 \Vh(n,g_s)
 ={(2\pi)^{\half n(n+1)}\over G_2(n+2)} V(n,g_s).\label{jtqh3Sep12}
\end{align}
Based on this observation, we define the ungauged partition function for
the lens space theory as follows:
\begin{align}
 \Zh_{\text{lens}}(N_1,N_2)_k
 &:=
 i^{-{\kappa\over 2}(N_1^2+N_2^2)}
 {(2\pi)^{\half N_1(N_1+1)+\half N_2(N_2+1)}\over G_2(N_1+2)G_2(N_2+2)}
 \int
 \prod_{j=1}^{N_1} {d\mu_j\over 2\pi}
 \prod_{a=1}^{N_2}{d\nu_a\over 2\pi}
 \notag\\
 &\qquad\qquad\times 
 \Delta_{\rm sh}(\mu)^2
 \Delta_{\rm sh}(\nu)^2
 \Delta_{\rm ch}(\mu,\nu)^2
 e^{-{1\over 2g_s}(\sum_j \mu_j^2+\sum_a \nu_a^2)}
 \\
 &=
 {(2\pi)^{\half N_1(N_1+1)+\half N_2(N_2+1)}
 \over  G_2(N_1+1) G_2(N_2+1) }
 Z_{\text{lens}}(N_1,N_2),
\end{align}
where we used the relation $G_2(n+2)=n!\, G_2(n+1)$.  The weak coupling
limit ($g_s\to 0,k\to\infty$) of this is
\begin{align}
 \Zh_{\text{lens}}(N_1,N_2)_{k\to \infty}
 =
  i^{-{\kappa\over 2}(N_1^2+N_2^2)} 
 2^{2N_1N_2}(2\pi g_s)^{N_1^2+N_2^2\over 2},
 \label{nimf2Nov12}
\end{align}
which does not involve $G_2$ or $N_2!$.  In a similar manner, we define
the ungauged partition function for the ABJ theory by
\begin{align}
 \Zh_{\text{ABJ}}(N_1,N_2)_k
 &:=
 i^{-{\kappa\over 2}(N_1^2-N_2^2)}
 {(2\pi)^{\half N_1(N_1+1)+\half N_2(N_2+1)}\over G_2(N_1+2)G_2(N_2+2)}
 \int
 \prod_{j=1}^{N_1} {d\mu_j\over 2\pi}
 \prod_{a=1}^{N_2}{d\nu_a\over 2\pi}
 \notag \\
 & \qquad\qquad \times 
 \Delta_{\rm sh}(\mu)^2
 \Delta_{\rm sh}(\nu)^2
 \Delta_{\rm ch}(\mu,\nu)^{-2}
 e^{-{1\over 2g_s}(\sum_j \mu_j^2-\sum_a \nu_a^2)}
 \\
 &=
 {(2\pi)^{\half N_1(N_1+1)+\half N_2(N_2+1)} \over  G_2(N_1+1) G_2(N_2+1) }
 Z_{\text{ABJ}}(N_1,N_2).
\end{align}
The weak coupling limit of this is
\begin{align}
 \Zh_{\text{ABJ}}(N_1,N_2)_{k\to\infty }
 =
 i^{-{\kappa\over 2}(N_1^2-N_2^2)}
 2^{-2N_1N_2}(2\pi g_s)^{N_1^2+N_2^2\over 2}.\label{nimv2Nov12}
\end{align}
By comparing \eqref{nimf2Nov12} and \eqref{nimv2Nov12}, we find that the
tree level partition functions are related simply as
\begin{align}
 \Zh_{\text{lens,tree}}(N_1,-N_2)_k=\Zh_{\text{ABJ,tree}}(N_1,N_2)_k.
\end{align}
Therefore, including the perturbative part, we expect that the full
partition functions satisfy
\begin{align}
 \Zh_{\text{lens}}(N_1,-N_2)_k=\Zh_{\text{ABJ}}(N_1,N_2)_k.\label{fjjc11Dec12}
\end{align}
We will see that this indeed holds in explicit examples.

In terms of $\Zh_{\text{lens}}$, our result \eqref{gpmz3Sep12} for the
lens space partition function can then be written as
\begin{align}
 \Zh_{\text{lens}}(N_1,N_2)_k
 &=
 i^{-{\kappa\over 2}(N_1^2+N_2^2)}
 (2\pi)^{N_1^2+N_2^2\over 2}g_s^{N_1+N_2\over 2}
\notag\\
 &\qquad\qquad
 \times
 q^{-{1\over 6}N(N^2-1)}(1-q)^{\half N(N-1)}
 B(N_1+N_2,N_1,N_2)\,S(N_1,N_2),\label{mnyq3Sep12}
\end{align}
where we defined
\begin{align}
 B(l,m,n):= {G_2(l+1;q)\over G_2(m+1)G_2(n+1)}.
\end{align}
Recall that $G_2(z)$ has zeros at $z=0,-1,-2,\dots$. Therefore,
$B(l,m,n)$ for $l,m,n\in\bbZ$ is finite if $m,n\ge 0$ but can be divergent if $m\le 0$ or
$n\le 0$.

In going from the lens space matrix model to the ABJ matrix model, we
flipped the sign of the quadratic term for $\nu_a$.  However, we could
have flipped the sign of the quadratic term for $\mu_j$.  This implies a
simple relation between $\Zh_{\text{lens}}(N_1,-N_2)_{k}$ and
$\Zh_{\text{lens}}(N_2,-N_1)_{k}$.  The relation is
\begin{align}
 \Zh_{\text{lens}}(N_1,-N_2)_k
 =\Zh_{\text{lens}}(N_2,-N_1)_{-k}.\label{lcqb22Sep12}
\end{align}
Here we have $-k$ on the right hand side because flipping the sign of
the quadratic term in $\mu_j$, not $\nu_a$, will change the sign of
$g_s\to -g_s$ in perturbative expansion.  In view of the relation
\eqref{fjjc11Dec12}, this is nothing but \eqref{N1N2swap}.

\subsection{Analytic continuation}
\label{Appendix_AC}

We would like to analytically continue $\Zh_{\text{lens}}(N_1,N_2)_k$ in
$N_2$.  The explicit expression for $\Zh_{\text{lens}}(N_1,N_2)_k$ is
given by \eqref{mnyq3Sep12}.  In particular, we are interested in
continuing $N_2$ to a negative integer $-N_2'$ where $N_2'\in
\bbZ_{>0}$.  However, this is not so simple because Barnes $G_2(z)$
vanishes for negative integral $z$ and hence $B(N_1+N_2,N_1,N_2)$ in
\eqref{mnyq3Sep12} diverges at $N_2=-N_2'$.  To deal with this
situation, let us analytically continue $N_2$ to
\begin{align}
 N_2=-N_2'+\epsilon,\qquad N_2'\in\bbZ_{>0},\qquad |\epsilon|\ll 1
\end{align}
and send $\epsilon\to 0$ at the end of the computation.  Using the
behavior of $G_2(z;q),G_2(z)$ near negative integral $z$ given in
\eqref{ih4Nov12} and \eqref{il4Nov12}, one can show that $B(N_1+N_2,N_1,N_2)$
diverges as $\epsilon\to 0$ as
\begin{align}
 &B(N_1+N_2,N_1,N_2)=B(N_1-N_2'+\epsilon,N_1,-N_2'+\epsilon)
\notag\\[2ex]
 &\qquad
 =
 \begin{cases}
   (-1)^{\half N_2'(N_2'-1)}
B(N_1-N_2',N_1,N_2')
%{G_2(N_1-N_2'+1;q)\over G_2(N_1+1)G_2(N_2'+1)}
  \,\epsilon^{-N_2'}
  & (N_2'\le N_1),\\[2ex]
   (-1)^{N_1N_2'+\half N_1(N_1+1)}q^{{1\over 6}(N_1-N_2')((N_1-N_2')^2-1)}&\\
  \qquad\qquad \times
  (1-q)^{N_1-N_2'}g_s^{-N_1+N_2'}
  B(N_2'-N_1,N_1,N_2')
%  {G_2(N_2'-N_1+1;q)\over G_2(N_1+1)G_2(N_2'+1)}
  \,\epsilon^{-N_1}
  & (N_1\le N_2'),
 \end{cases}
 \label{iamq4Sep12}
\end{align}
where we only kept the leading term.
Therefore, in order for the entire $\Zh_{\text{lens}}$ to remain finite
as $\epsilon\to 0$, the function $S(N_1,-N_2'+\epsilon)$ should vanish
as
\begin{align}
 S(N_1,-N_2'+\epsilon)
 \sim\begin{cases}
   \epsilon^{N_2'}& (N_2'\le N_1)\\
   \epsilon^{N_1} & (N_1\le N_2')
     \end{cases}
 ~~
 =\epsilon^{\min(N_1,N_2')}.\label{mvqa4Sep12}
\end{align}

In the following, we will explicitly carry out analytic continuation of
$S(N_1,N_2)$ and find that it indeed behaves as \eqref{mvqa4Sep12}.  We
will begin with  the simple cases with $N_1=1,2$ to get the hang of
it, and then move on to the general $N_1$ case.

\subsubsection{$\boldsymbol{N_1=1}$}
\label{Appendix_N1equalto1}

The simplest case is $N_1=1$, for which \eqref{msmi31Aug12} gives
\begin{align}
 S(1,N_2)
 &= \sum_{C=1}^{N_2+1}
 \prod_{C<a}{q^{C}+q^{a}\over q^{C}-q^{a}}
 \prod_{a<C}{q^{a}+q^{C}\over q^{a}-q^{C}}
 = \sum_{C=1}^{N_2+1}
 \prod_{j=1}^{N_2-C+1}{1+q^j\over 1-q^j}
 \prod_{j=1}^{C-1}{1+q^j\over 1-q^j}\\
 &= 
 \sum_{C=1}^{N_2+1}
 {(-q)_{N_2-C+1} \over (q)_{N_2-C+1}}
 {(-q)_{C-1} \over (q)_{C-1}}
 =\sum_{n=0}^{N_2}
 {(-q)_{N_2-n} \over (q)_{N_2-n}}
 {(-q)_{n} \over (q)_{n}},
 \label{gjqa20Sep12}
\end{align}
where $n=C-1$. $(a)_n$ is the $q$-Pochhammer symbol defined in Appendix \ref{Appendix_q-Analogs}\@.  We
want to analytically continue this expression in $N_2$.  The explicit
$N_2$ dependence of the sum range seems to be an obstacle, but it can be
circumvented by the following observation: as a function of $z$, $(q)_z$
has poles of order 1 at $z\in \bbZ_{<0}$, while $(-q)_z$ has no poles.
Therefore, the summand in \eqref{gjqa20Sep12} vanishes unless $0\le n\le
N_1$, and we can actually extend the range of summation as
\begin{align}
 S(1,N_2)
 &= \sum_{n=0}^{\infty}
 {(-q)_{N_2-n} \over (q)_{N_2-n}}
 {(-q)_{n} \over (q)_{n}}.\label{lnuo31Aug12}
\end{align}
This expression can be analytically continued to complex $N_2$,
including negative integers.\footnote{We did not make $n$ to run over
the entire $\bbZ$ because it would give $S=0$. Namely, including $n\in
\bbZ_{<0}$ would exactly cancel the contribution from $n\in \bbZ_{\ge
0}$. Showing this requires to regularize the sum, for example, by $n\to
n+\eta$ for $n\in \bbZ_{<0}$ with $\eta\to 0$.}

We can rewrite \eqref{lnuo31Aug12} in different forms which we will find
more convenient.  First, using \eqref{jwdu20Sep12} and
\eqref{jzmp20Sep12}, one can show that
\begin{align}
 S(1,N_2) = \beta(1,N_2)\,\Phi(1,N_2),\label{mckp3Sep12}
\end{align}
where
\begin{align}
 \beta(1,N_2):= {(-q)_{N_2}\over (q)_{N_2}},\qquad
 \Phi(1,N_2):=\sum_{n=0}^\infty
 {(-1)^n}{(q^{-N_2})_n (-q)_n\over (-q^{-N_2})_n (q)_n}
 =
 \, _2\phi_1\left({q^{-N_2},-q\atop -q^{-N_2}} ; q,-1\right).\label{evuj11Sep12}
\end{align}
This expression is useful because the relation to $q$-hypergeometric
function is manifest.  The $q$-hypergeometric function $_2\phi_1$ is defined
in Appendix \ref{Appendix_q-Analogs}\@. In addition, this way of writing $S$ is useful because
it splits it into $\beta$ which vanishes for negative integral
$N_2\in\bbZ_{<0}$ and $\Phi$ which is finite for all $N_2\in\bbZ$.  It
is easy to see that the first factor $\beta$ vanishes for negative
$N_2=-N_2'\in\bbZ_{<0}$:
\begin{align}
 \beta(1,-N_2')
 = {(-q)_{-N_2'}\over (q)_{-N_2'}}
 = {(q^{1-N_2'})_{N_2'}\over (-q^{1-N_2'})_{N_2'}}
 = {(1-q^{1-N_2'})\cdots (1-q^0)\over 
 (1+q^{1-N_2'})\cdots (1+q^0)}=0.
\label{kjhl20Sep12}
\end{align}
However, we are actually setting $N_2=-N_2'+\epsilon$ and we have to
keep track of how fast this vanishes as $\epsilon\to 0$.
$\beta(1,-N_2'+\epsilon)$ involves ${(\pm q)_{-N_2'+\epsilon}}$ which,
using \eqref{hcey4Sep12} with $z=-1+\epsilon$ and \eqref{hcit4Sep12},
can be rewritten as
\begin{align}
 (\pm q)_{-N_2'+\epsilon}
 &=(\mp 1)^{N_2'-1}q^{-\half N_2'(N_2'+1)}{(\pm q)_{-1+\epsilon}\over (\pm q)_{N_2'-1}}.\label{msfw4Sep12}
\end{align}
The behavior of $(\pm q)_{-1+\epsilon}$ can be seen,
using the definition \eqref{maux20Sep12}, as follows:
\begin{align}
 (q)_{-1+\epsilon}
 &
 ={(1-q)(1-q^2)\cdots\over (1-q^\epsilon)(1-q^{1+\epsilon})\cdots}
 =-{1\over \epsilon\ln q},
 \quad
 (-q)_{-1+\epsilon}
 ={(1+q)(1+q^2)\cdots\over (1+q^\epsilon)(1+q^{1+\epsilon})\cdots}
 ={1\over 2},
 \label{mbuy20Sep12}
\end{align}
where we kept only leading terms. We will do this kind of manipulation
to extract $\epsilon\to 0$ behavior over and over again below, but we
will not present the details henceforth.  So, the behavior of
$\beta(1,N_2)$ near integral $N_2$ is
\begin{align}
 \beta(1,N_2+\epsilon)=
\begin{cases}
 {(-q)_{N_2}\over (q)_{N_2}}
 & (N_2>0),
 \\
 (-1)^{N_2'} {\epsilon \ln q\over 2} {(q)_{N_2'-1}\over (-q)_{N_2'-1}}
 & (N_2=-N_2'<0).
\end{cases} 
\label{gnlw21Sep12}
\end{align}
The $\cO(\epsilon)$ behavior for $N_2<0$ is the correct one to cancel
the divergence of $B$ that we saw in
\eqref{iamq4Sep12}, \eqref{mvqa4Sep12}.
On the other hand, the second factor $\Phi$ in \eqref{mckp3Sep12} is
finite for all $N_2\in\bbZ$.  For $N_2>0$, $(q^{-N_2})_n$
becomes zero for $n\ge N_2+1$ and the sum reduces to a finite sum.  For
$N_2=-N_2'<0$, the sum $(q^{-N_2})_n=(q^{N_2'})_n$ is
non-vanishing for all $n\ge 0$.

There is another useful expression for $S(1,N_2)$.  Using $q$-Pochhammer
formulas, we can show that
\begin{align}
 S(1,N_2)=\gamma(1,N_2)\,\Psi(1,N_2),\label{jyes7Nov12}
\end{align}
where
\begin{align}
 \gamma(1,N_2):= {(-q)_{N_2}(-q)_{-N_2-1}\over (q)_{N_2}(q)_{-N_2-1}},\qquad
 \Psi(1,N_2):=\sum_{s=0}^\infty
 (-1)^s{(q^{s+1})_{-N_2-1}\over (-q^{s+1})_{-N_2-1}}
\label{gowu21Sep12}
\end{align}
and we relabeled $n\to s$.  This expression is useful because some
symmetries are more manifest, as we will see later in the $N_1\ge 2$
cases.  At the same time, however, $\Psi$ is slightly harder to deal
with for $N_2>0$ than $\Phi$, because $(q^{s+1})_{-N_2-1}={1\over
(q^{s-N_2})_{N_2+1}}$ can diverge.  So, in this way of writing $S$, we
should introduce $\epsilon$ even for $N_2>0$ and set $N_2\to
N_2+\epsilon$.
Just as we did for $\beta$, we can evaluate $\gamma(1,N_2)$ near
integral $N_2$ and the result is
\begin{align}
 \gamma(1,N_2+\epsilon)
 =
 (-1)^{N_2}\,{\epsilon \ln q\over 2} \qquad
 \qquad \text{for all $N_2\in\bbZ$.}
\label{gowz21Sep12}
\end{align}
For $N_2<0$, this just cancels the $\epsilon^{-1}$ divergence from $B$
given in \eqref{iamq4Sep12}, while $\Psi$ is finite.  For $N_2>0$, for
which $B$ is finite, the $\epsilon$ coming from \eqref{gowz21Sep12} is
canceled by $\Psi$ which goes as $\epsilon^{-1}$ in this case.  In more
detail, for $N_2>0$, it is only the $0\le s\le N_2$ terms in $\Psi$ that
behave as $\epsilon^{-1}$ and cancel against $\gamma\sim \epsilon$,
whereas the $s>N_2$ terms are finite and vanish when multiplied by
$\gamma\sim\epsilon$.  This is a complicated way to say that, in the sum
\eqref{lnuo31Aug12}, only $0\le s\le N_2$ terms contribute.

Introduction of all these quantities may seem unnecessary complication,
but this will become useful in more general $N_1\ge 2$ cases discussed
below. How various quantities behave as $\epsilon\to 0$ is summarized in
Table \ref{table:epsilonBehavior1}.

\begin{table}
 \begin{center}
 \begin{tabular}{|c||c|cc|cc|c|c|}
  \hline
  & $B$ & $\beta$ & $\Phi$ & $\gamma$ & $\Psi$ & $S=\beta \Phi=\gamma\Psi$ & $\Zh\propto B S$\\
  \hline\hline
  $N_2>0$ & finite & finite & finite & $\epsilon$ & $\epsilon^{-1}$ & finite & finite\\
  \hline
  $N_2<0$ & $\epsilon^{-1}$ & $\epsilon$ & finite & $\epsilon$ & finite & $\epsilon$ & finite\\
  \hline
 \end{tabular}
 \end{center}
 \caption{\sl The $\epsilon\to 0$ behavior of various quantities for
 $N_1=1$.  Although $B$ and
 $S=\beta\Phi=\gamma\Psi$ can be individually singular, the partition
 function $\Zh\propto BS$ is always finite.  \label{table:epsilonBehavior1}}
\end{table}

Now we are ready to present the expression for the analytically
continued partition function $\Zh_{\text{lens}}$ for $N_1=1$ and
$N_2=-N_2'<0$. Combining \eqref{gowu21Sep12} and \eqref{gowz21Sep12},
and using \eqref{iamq4Sep12}, we obtain the expression for the ABJ
partition function
$\Zh_{\text{ABJ}}(1,N_2')_k=\Zh_{\text{lens}}(1,-N_2')_k$:
\begin{align}
 \Zh_{\text{ABJ}}(1,N_2')_k
 &=
 i^{-{\kappa\over 2}(1+N_2'^2)}
 (2\pi)^{1+N_2'^2\over 2}g_s^{1+N_2'\over 2}
 (1-q)^{(N_2'-1)(N_2'-2)\over 2}
 {G_2(N_2';q)\over 2G_2(N_2'+1)}
 \Psi(1,-N_2'),
\label{gwsv10Sep12}
\end{align}
where
\begin{align}
 \Psi(1,-N_2')&=\sum_{s=0}^\infty
 (-1)^s{(q^{s+1})_{N_2'-1}\over (-q^{s+1})_{N_2'-1}}.
\end{align}

\subsubsection{$\boldsymbol{N_1=2}$}
\label{Appendix_N1equalto2}

For $N_1=2$, the general formula \eqref{msmi31Aug12} gives the
following expression for $S$:
\begin{align}
 S(2,N_2)&=
 \sum_{1\le C_1<C_2\le N_2+2}
 \prod_{a=C_1+1}^{C_2-1}{q^{C_1}+q^{a}\over q^{C_1}-q^a}
 \prod_{a=C_2+1}^{N_2+2}{q^{C_1}+q^{a}\over q^{C_1}-q^a}
 \prod_{a=C_2+1}^{N_2+2}{q^{C_2}+q^{a}\over q^{C_2}-q^a}\notag\\
 &\qquad\qquad\qquad\qquad
 \times
 \prod_{a=1}^{C_1-1}{q^a+q^{C_1}\over q^a-q^{C_1}}
 \prod_{a=1}^{C_1-1}{q^a+q^{C_2}\over q^a-q^{C_2}}
 \prod_{C_1+1}^{C_2-1}{q^a+q^{C_1}\over q^a-q^{C_1}}\\
 &=
 \sum_{1\le C_1<C_2\le N_2+2}
 {(-q)_{C_2-C_1-1}\over (q)_{C_2-C_1-1}}
 {(-q^{C_2-C_1+1})_{N_2-C_2+2}\over (q^{C_2-C_1+1})_{N_2-C_2+2}}
 {(-q)_{N_2-C_2+2}\over (q)_{N_2-C_2+2}}
 \notag\\
 &\qquad\qquad\qquad\qquad
 \times
 {(-q)_{C_1-1}\over (q)_{C_1-1}}
 {(-q^{C_2-C_1+1})_{C_1-1}\over (q^{C_2-C_1+1})_{C_1-1}}
 {(-q)_{C_2-C_1-1}\over (q)_{C_2-C_1-1}}.\label{hndk21Sep12}
\end{align}
Just as we did for the $N_1=1$ case, we want to analytically continue
this expression by eliminating the explicit $N_2$ dependence of the sum
range by extending it.  However, this turns out to be a non-trivial
issue and, in particular, the way to do it is not unique.  Before
discussing it, let us first consider rewriting $S$ in different forms.

First, just as in the $N_1=1$ case, we can rewrite $S$ in a form closely
related to $q$-hypergeometric
functions.  Namely,
\begin{align}
 S(2,N_2)&=\beta(2,N_2)\,\Phi(2,N_2),\label{kgmb7Nov12}
\end{align}
where
\begin{align}
\begin{split}
 \beta(2,N_2)&={(-q)_{N_2}(-q^2)_{N_2}\over (q)_{N_2}(q^2)_{N_2}},\\
 \Phi(2,N_2)&=\sum_{n_1,n_2}
 (-1)^{n_2}
 {(-q)_{n_1} (q^{-N_2-1})_{n_1}\over (q)_{n_1} (-q^{-N_2-1})_{n_1}}
 {(-q)_{n_2}^2 (q^2)_{n_2}^2\over (q)_{n_2}^2 (-q^2)_{n_2}^2}
 {(q^{-N_2})_{{n_1}+{n_2}}(-q^2)_{{n_1}+{n_2}}\over (-q^{-N_2})_{{n_1}+{n_2}}(q^2)_{{n_1}+{n_2}}}
\end{split}
\label{gyiq5Sep12}
\end{align}
and $C_1-1=n_1,C_2-C_1-1=n_2$.  The original range of summation
corresponds to $n_1\ge 0,n_2\ge 0,n_1+n_2\le N_2$, but we did not
specify the range here for the reason mentioned above.  This expression
is the analogue of the $N_1=1$ relation \eqref{mckp3Sep12}; $\beta$
diverges for $N_2<0$ while $\Phi$ is finite for both $N_2>0$ and
$N_2<0$.  $\Phi$ has the same form as the double $q$-hypergeometric
function defined in \cite{Gasper-Rahman}, if the summation were over
$n_1,n_2\ge 0$.

The second expression for $S$ is 
\begin{align}
 S(2,N_2)&=\gamma(2,N_2)\,\Psi(2,N_2),
\end{align}
where
\begin{align}
 \gamma(2,N_2)&=
 -
 {(-q)_1^2\over (q)_1^2}
 {(-q)_{N_2}(-q^2)_{N_2}\over (q)_{N_2}(q^2)_{N_2}}
 {(-q)_{-N_2-2}(-q^2)_{-N_2-2}\over (q)_{-N_2-2}(q^2)_{-N_2-2}},\notag\\
 \Psi(2,N_2)&=\sum_{s_1,s_2} (-1)^{s_1+s_2}
 {(q^{s_1+1})_{-N_2-2} \over (-q^{s_1+1})_{-N_2-2}}
 {(q^{s_2+1})_{-N_2-2} \over (-q^{s_2+1})_{-N_2-2}}
 {(q^{s_2-s_1})_1^2  \over (-q^{s_2-s_1})_1^2}\label{hnsz21Sep12}
\end{align}
and $s_1=C_1-1,s_2=C_2-1$. This expression is the analogue of
\eqref{jyes7Nov12}.  The original range of summation
corresponds to $0\le s_1<s_2\le N_2+1$.

Now let us discuss the issue of the sum range. For the purpose of
studying when the summand vanishes, the $\beta\Phi$ expression
\eqref{kgmb7Nov12} is convenient, because $\beta$ just cancels the
divergence of $B$ while $\Phi$ is always finite.  So, all we need to
know is when the summand in $\Phi$ vanishes.  Note that, when
regularized, $(q^m)_n$ with $m,n\in\bbZ$ has the following behavior:
\begin{align}
\begin{split}
 n>0: & \quad
 (q^m)_n = (1-q^m)\cdots (1-q^{m+n-1})
 = \begin{cases}
       \cO(\epsilon) & m\le 0 ~\text{and} ~ m+n-1\ge 0,\\
       \cO(1) &\text{otherwise},
      \end{cases}
 \\
 n<0: & \quad
 (q^m)_n = {1\over (q^{m+n})_{-n}}
 ={1\over (1-q^{m+n})\cdots (1-q^{m-1})}\\
&\qquad \qquad \qquad \qquad \quad 
 = \begin{cases}
       \cO(\epsilon^{-1}) & m+n\le 0 ~\text{and} ~ m-1\ge 0,\\
       \cO(1) &\text{otherwise}.
   \end{cases}
\end{split}\label{hdrp8Nov12}
\end{align}
Here, regularizing $(q^m)_n$ means to replace $N_2$ entering $m,n$ by
$N_2+\epsilon$.  Furthermore, when $n_1,n_2<0$, we must regularize the
summand in \eqref{gyiq5Sep12} by setting $n_1\to n_1+\eta$, $n_2\to n_2+\eta$ with
$\eta\to 0$. In this case, we must replace $\cO(\epsilon)$ in
\eqref{hdrp8Nov12} by $\cO(\epsilon,\eta)$ and
$\cO(\epsilon^{-1})$ by $\cO(\epsilon^{-1},\eta^{-1})$.
Using this, it is straightforward to determine the range of $(n_1,n_2)$
for which the summand in $\Phi$ remains non-vanishing after setting
$\epsilon,\eta\to 0$.

\begin{figure}[h!]
\begin{center}
 \includegraphics[height=2.9in]{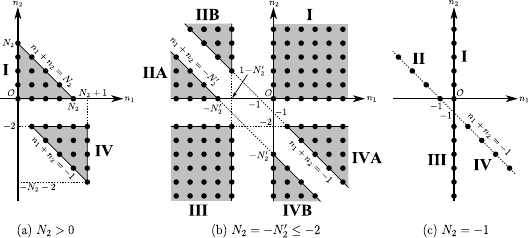}
 \caption{\sl The regions that can contribute to $\Phi(1,N_2)$.  (a), (b): For $(n_1,n_2)\in\bbZ^2$ in the shaded regions (denoted
 by dots), the summand in $\Phi(1,N_2)$ in \eqref{gyiq5Sep12} is $\cO(1)$.
 Outside the shaded regions, the summand is $\cO(\epsilon,\eta)$ and
 vanishes as $\epsilon,\eta\to 0$. (c): the $N_2=-1$ case  special and
the summand is non-vanishing only on the dots.
 \label{nonvanishing_n1n2} }
\end{center}
\end{figure}

In Figure \ref{nonvanishing_n1n2}, we described the regions in the
$(n_1,n_2)$ plane in which the summand appearing in $\Phi(1,N_2)$ is
non-vanishing.  Figure \ref{nonvanishing_n1n2}(a) shows that, for
$N_2>0$, the summand is non-vanishing in the original range of
summation, $n_1\ge 0,n_2\ge 0,n_1+n_2\le N_2$ (region I), as it should
be.  We would like to extend the range in order to eliminate the $N_2$
dependence and thereby analytically continue $\Phi(1,N_2)$ to negative
$N_2$.  The requirements for the extension are
\begin{enumerate}
 \item[(i)] The range specification does not involve $N_2$,
 \item[(ii)] For $N_2>0$, it reproduces the original result
	    \eqref{hndk21Sep12}.
\end{enumerate}
Clearly, there are more than one ways to extend the range satisfying
these requirements.  One simple way would be to take $n_1\ge0,n_2\ge 0$
as the extended range.  For $N_2>0$, this reduces to region I and
reproduces the original result, while for $N_2<0$ this sums over region
I in Figure \ref{nonvanishing_n1n2}(b). (We consider $N_2\le -2$, since
$N_2=-1$ is rather exceptional as one can see in Figure
\ref{nonvanishing_n1n2}(c). The latter case will be discussed later.)
Another possible extension is $n_2\ge 0$.  This also reproduces the
original result for $N_2>0$, but for $N_2<0$ this sums over not only
regions I but also IIA and IIB\@.

Therefore, the way to analytically continue $\Phi(1,N_2)$ is ambiguous
and, mathematically, any such choices are good (ignoring the fact that
the sum may not be convergent and is only formal). Namely, the data for
discrete $N_2\in\bbZ_{>0}$ is not enough to uniquely determine the
analytic continuation for all $N_2\in\bbC$. Additional input comes from
the physical requirement that it reproduce the known ABJ results
for $N_2<0$.  Furthermore, for $N_2=-1$, $\Zh_{\text{lens}}(2,-1)_k$ is
expected to be related to $\Zh_{\text{lens}}(1,-2)_{-k}$ by the relation
\eqref{lcqb22Sep12}.

Here we simply present the prescription which satisfies these physical
requirements.  The explicit checks are done in the main text where it is
shown that its perturbative expansions agree with the known ABJ result
and, when exact non-perturbative expressions for the ABJ matrix integral
are known, it reproduces them.  Moreover, the fact that the prescription
reproduces the relation between $\Zh_{\text{lens}}(2,-1)_k$ and
$\Zh_{\text{lens}}(1,-2)_{-k}$ is shown for general $N_1$ below.
 
\begin{figure}[htbp]
\begin{center}
 \includegraphics[height=2.6in]{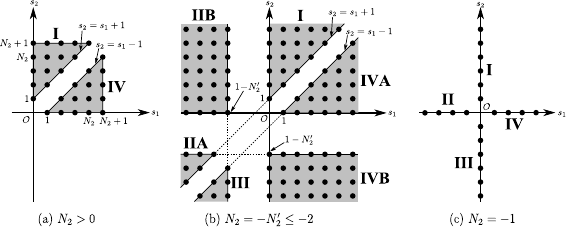} \caption{\sl The
 regions that can contribute to $\Phi(1,N_2)$.  These are the same as
 Figure \ref{nonvanishing_n1n2}, but plotted for $(s_1,s_2)$ instead.
 \label{nonvanishing_s1s2} }
\end{center}
\end{figure}

The key observation to arrive at such a prescription is that, as we can
see from Figure \ref{nonvanishing_n1n2}(a), the summand is non-vanishing
not only in the original region I but also in region IV\@.  The meaning
of this is easier to see in the $\gamma\Psi$ representation in terms of
$s_1,s_2$.  In Figure \ref{nonvanishing_s1s2}, we presented the same
diagram as Figure \ref{nonvanishing_n1n2} but on the $(s_1,s_2)$ plane.
As we can see from the Figure, the non-vanishing regions have the
symmetry
\begin{align}
 s_1\leftrightarrow s_2.\label{hnla21Sep12}
\end{align}
Actually, as we can immediately see from the explicit expression for
$\Psi$ given in \eqref{hnsz21Sep12}, this is a symmetry of the summand,
not just its non-vanishing regions.  Therefore, it is natural to
relax the ordering constraint $s_1<s_2$ in the original range and
sum over both regions I and IV, after dividing by $2$.  If
$s_1=s_2$, the summand in \eqref{hnsz21Sep12} automatically vanishes.
Namely, we can write $\Psi$ as
\begin{align}
 \Psi(2,N_2)={1\over 2}\sum_{s_1,s_2=0}^\infty  (-1)^{s_1+s_2}
 {(q^{s_1+1})_{-N_2-2} \over (-q^{s_1+1})_{-N_2-2}}
 {(q^{s_2+1})_{-N_2-2} \over (-q^{s_2+1})_{-N_2-2}}
 {(q^{s_2-s_1})_1^2  \over (-q^{s_2-s_1})_1^2}.\label{glud9Nov12}
\end{align}
Here we have extended the sum range so that $s_1,s_2$ run to infinity,
which is harmless in the $N_2>0$ case.

Our prescription is that we use the expression \eqref{glud9Nov12} even
for $N_2=-N_2'<0$.  As we can see from Figure
\ref{nonvanishing_s1s2}(b), this sums over regions I and IVA\@.  As we
have been emphasizing, it is by no means clear at this point that this
is the right prescription. The justification is given in the main text
where it is shown that this is consistent with all known results.  One
can also show that the other possible prescriptions, such as $n_1\ge
0,n_2\ge 0$, which covers region I, and $n_2\ge 0$, which covers regions
I, IIA and IIB, would not reproduce the known results and hence are not
correct.

If we set $N_2\to N_2+\epsilon$, the behavior of $\gamma$ is
\begin{align}
 \gamma(2,N_2+\epsilon)
 =
\left({\epsilon\ln q\over 2}\right)^2 
 \qquad\text{for all $N_2\in\bbZ$.}
 \label{irrq21Sep12}
\end{align}
Substituting this and \eqref{iamq4Sep12} into \eqref{mnyq3Sep12}, we
finally obtain the expression for the ABJ partition function
$\Zh_{\text{ABJ}}(2,N_2')_k=\Zh_{\text{lens}}(2,-N_2')_k$:
\begin{align}
 \Zh_{\text{ABJ}}(2,N_2')_k
 &=i^{-{\kappa\over 2}N_2'^2}
 (2\pi)^{2+{N_2'^2\over 2}}g_s^{1+{N_2'\over 2}}
 (1-q)^{\half(N_2'-2)(N_2'-3)}
 {G_2(N_2'-1;q)\over 4G_2(N_2'+1)}
 \Psi(2,-N_2'),
 \label{jekb21Sep12}
\end{align}
where it is assumed that $N_2'\ge 2$ and $\Psi(1,-N_2')$ is given simply by setting $N_2=-N_2'$ in \eqref{glud9Nov12}:
\begin{align}
  \Psi(2,-N_2')={1\over 2}\sum_{s_1,s_2=0}^\infty  (-1)^{s_1+s_2}
 {(q^{s_1+1})_{N_2'-2} \over (-q^{s_1+1})_{N_2'-2}}
 {(q^{s_2+1})_{N_2'-2} \over (-q^{s_2+1})_{N_2'-2}}
 {(q^{s_2-s_1})_1^2  \over (-q^{s_2-s_1})_1^2}.
\end{align}

The above formula is valid for $N_2'\ge 2$ but not for $N_2'=1$.  This
case is important, because $(N_1,N_2')=(2,1)$ is related to
$(N_1,N_2')=(1,2)$ by \eqref{lcqb22Sep12} and therefore the summation
over two variables $s_1,s_2$ should truncate to a sum with one variable;
this provides a further check of our prescription.  We will discuss this
more generally below, where we discuss general $N_1$.

\subsubsection{General $\boldsymbol{N_1}$}
\label{Appendix_GenN1}

With the $N_1=1,2$ cases understood, the prescription for general $N_1$
is straightforward to establish, although computations get cumbersome.
Much as in the $N_1=1,2$ cases, the general expression for $S$ in
\eqref{msmi31Aug12} can be rewritten in the following form:
\begin{align}
 S(N_1,N_2)
 &=
 \sum_{1\le C_1<\cdots<C_{N_1}\le N}
 \prod_{j=1}^{N_1} \Biggl\{\Biggl[\prod_{k=j}^{N_1-1}\prod_{a=C_k+1}^{C_{k+1}-1}{q^{C_j}+q^a\over q^{C_j}-q^a}\Biggr]
 \prod_{a=C_{N_1}+1}^{N}{q^{C_j}+q^a\over q^{C_j}-q^a}\notag\\
 &\qquad\qquad\qquad\qquad\qquad
 \times
 \prod_{a=1}^{C_1-1}{q^a+q^{C_j}\over q^a-q^{C_j}}
 \Biggl[\prod_{k=1}^{j-1}\prod_{a=C_k+1}^{C_{k+1}-1}{q^a+q^{C_j}\over q^a-q^{C_j}}\Biggr]
 \Biggr\}
 \notag\\
 &=
 \sum_{1\le C_1<\cdots<C_{N_1}\le N}
 \prod_{j=1}^{N_1} \Biggl\{\Biggl[\prod_{k=j}^{N_1-1}{(-q^{C_k-C_j+1})_{C_{k+1}-C_k-1}\over (q^{C_k-C_j+1})_{C_{k+1}-C_k-1}}\Biggr]
 {(-q^{C_{N_1}-C_j+1})_{N-C_{N_1}}\over (q^{C_{N_1}-C_j+1})_{N-C_{N_1}}}\notag\\
 &\qquad\qquad\qquad\qquad
 \times
 {(-q^{C_j-C_1+1})_{C_1-1}\over (q^{C_j-C_1+1})_{C_1-1}}
 \Biggl[\prod_{k=1}^{j-1}{(-q^{C_j-C_{k+1}+1})_{C_{k+1}-C_k-1}\over (q^{C_j-C_{k+1}+1})_{C_{k+1}-C_k-1}}\Biggr]
 \Biggr\}.\label{mskn15Sep12}
\end{align}
In expressions such as this, it is understood that
$\sum_{l=a}^b ...=0$ and 
$\prod_{l=a}^b ...=1$ if $a>b$.

Again, we can rewrite this in the $\beta \Phi$ and $\gamma \Psi$
representations.
The $\beta\Phi$ representation is
\begin{align}
 &S(N_1,N_2)=\beta(N_1,N_2)\,\Phi(N_1,N_2),\label{fvrs7Sep12}\\
 &\beta(N_1,N_2):=\prod_{j=1}^{N_1}{(-q^{N_1-j+1})_{N_2}\over (q^{N_1-j+1})_{N_2}},\\
 &\Phi(N_1,N_2):=
 \sum_{n_1,\dots,n_{N_1}}
 (-1)^{\sum_{l=1}^{N_1}(N_1-l+1)n_l}\notag\\
 &\quad\times
 \left[\prod_{j=1}^{N_1-1}\prod_{k=j}^{N_1-1}
 {( q^{k-j+1})_{n^{}_{j+1,k}}(-q^{k-j+1})_{n^{}_{j+1,k+1}}^2( q^{k-j+1})_{n^{}_{j+2,k+1}} \over
  (-q^{k-j+1})_{n^{}_{j+1,k}}( q^{k-j+1})_{n^{}_{j+1,k+1}}^2(-q^{k-j+1})_{n^{}_{j+2,k+1}}}
 \right]
 \notag\\
 &\quad\times
 \left[\,\prod_{j=1}^{N_1}
 {( q^{N_1-j+1})_{n^{}_{j+1,N_1}}( q^{-N_1-N_2+j})_{n^{}_{1,j}} \over
  (-q^{N_1-j+1})_{n^{}_{j+1,N_1}}(-q^{-N_1-N_2+j})_{n^{}_{1,j}}}
 \right]
 \left[\prod_{k=0}^{N_1-1}
 {(-q^{k+1})_{n^{}_{1,k+1}}( q^{k+1})_{n^{}_{2,k+1}} \over
  ( q^{k+1})_{n^{}_{1,k+1}}(-q^{k+1})_{n^{}_{2,k+1}} }
 \right],\label{hapn6Sep12}
\end{align}
where we defined $n_1=C_1-1$, $n_j=C_j-C_{j-1}-1$ ($j=2,\dots,N_1$), and
$n_{a,b}:= \sum_{l=a}^b n_l$.  Furthermore, we define
$n_{N_1+1,b}=n_{a,N_1+1}=0$.  The original sum range $1\le
C_1<\dots<C_{N_1}\le N$ corresponds to $n_j\ge 0$ ($j=1,\dots,N_1$),
$n_1+\cdots+n_{N_1}\le N_2$, but we did not specify it in
\eqref{hapn6Sep12} for the same reason as in the $N_1=2$ case.
$\Phi$ has the form of the multi-variable generalization of $q$-hypergeometric
functions, discussed {\it e.g.}\ in \cite{Exton}.
When we analytically continue by $N_2\to -N_2'+\epsilon$,
$\beta(N_1,-N_2'+\epsilon)$ goes to zero, while $\Phi(N_1,-N_2')$ remains
finite.  The behavior of $\beta$ as $\epsilon\to 0$ is
\begin{align}
 &\beta(N_1,-N_2'+\epsilon)
 =\begin{cases}
    \left(-{\epsilon\log q\over 2}\right)^{N_2'}(-1)^{\half N_2'(N_2'-1)}
    \prod_{j=1}^{N_2'-1}{(q)_j\over (-q)_j}
    \prod_{j=N_1-N_2'}^{N_1-1}{(q)_j\over (-q)_j}
    & (N_2'\le N_1),
   \\[2ex]
    \left(-{\epsilon\log q\over 2}\right)^{N_1}
    (-1)^{N_1N_2'+\half N_1(N_1+1)}
    \prod_{j=1}^{N_1-1}{(q)_j\over (-q)_j}
    \prod_{j=N_2'-N_1}^{N_2'-1}{(q)_j\over (-q)_j}
    & (N_1\le N_2').
  \end{cases}
 \label{euun7Sep12}
\end{align}

On the other hand, the $\gamma \Psi$ representation is
\begin{align}
 S(N_1,N_2)&=\gamma(N_1,N_2)\, \Psi(N_1,N_2),\label{moge23Sep12}\\
 \gamma(N_1,N_2)&=
 (-1)^{\half N_1(N_1-1)}
 \prod_{j=1}^{N_1-1}{(-q)_j^2\over (q)_j^2}
 \prod_{j=1}^{N_1}
  {(-q^j)_{N_2}(-q^j)_{-N_1-N_2}\over (q^j)_{N_2}(q^j)_{-N_1-N_2}},
 \label{fxmk24Sep12}
 \\
 \Psi(N_1,N_2)&=
 {1\over N_1!}\sum_{s_1,\dots,s_{N_1}=0}^{\infty}
 (-1)^{s_1+\cdots+s_{N_1}}
 \prod_{j=1}^{N_1}{(q^{s_j+1})_{-N_1-N_2}\over (-q^{s_j+1})_{-N_1-N_2}}
 \prod_{1\le j<k\le N_1}
 {(q^{s_k-s_j})_1^2\over (-q^{s_k-s_j})_1^2},
  \label{fudw24Sep12}
\end{align}
where $s_j:= C_j-1, j=1,\dots,N_1$.  The original sum range
corresponds to $0\le s_1<\cdots<s_{N_1}\le N-1$.  However, because of
the $s_j\leftrightarrow s_k$ symmetry of this expression, we can forget
about the ordering constraints and let $s_j$ run freely, if one divides
the expression by $N_1!$, which we have already done above.  Furthermore, just
as in the $N_1=2$ case, we can safely remove the upper bound in the
summation for $N_2>0$.  Our prescription for analytic continuation to
$N_2<0$ is to use this same expression \eqref{fudw24Sep12}, by setting
$N_2=-N_2'+\epsilon$ with $\epsilon\to 0$.

The behavior of $\gamma(N_1,N_2)$ near integral $N_2$ can be shown to be
\begin{align}
 \gamma(N_1,N_2+\epsilon)
 =  (-1)^{N_1N_2+N_1} \left(-{\epsilon\ln q\over 2}\right)^{N_1}
 \qquad \text{for all $N_2\in\bbZ$.}
\label{fxom24Sep12}
\end{align}
By substituting \eqref{moge23Sep12} and \eqref{iamq4Sep12} into
\eqref{mnyq3Sep12}, we obtain the expression for the ABJ partition
function $\Zh_{\text{ABJ}}(N_1,N_2')_k=\Zh_{\text{lens}}(N_1,-N_2')_k$:
\begin{align}
 \Zh_{\text{ABJ}}(N_1,N_2')_k
 &=
 i^{-{\kappa\over 2}(N_1^2+N_2'^2)}
 (-1)^{\half N_1(N_1-1)}
 2^{-N_1}(2\pi)^{N_1^2+N_2'^2\over 2}
 g_s^{N_1+N_2'\over 2}
 \notag\\
 &\qquad\qquad\times
 (1-q)^{\half(N_2'-N_1)(N_2'-N_1-1)}
 B(N_2'-N_1,N_1,N_2)
 \Psi(N_1,-N_2')
 \notag\\ 
 &=
 i^{-{\kappa\over 2}(N_1^2+N_2'^2)}
 (-1)^{\half N_1(N_1-1)}
 2^{-N_1}(2\pi)^{N_1^2+N_2'^2\over 2}
 g_s^{N_1+N_2'\over 2}
 \notag\\
 &\qquad\qquad\qquad
 \times
 {\prod_{j=1}^{N_2'-N_1-1}(q)_j\over G_2(N_1+1)G_2(N_2'+1)}
 \Psi(N_1,-N_2'),
 \label{nijj23Sep12}
\end{align}
where we assumed that $N_2'\ge N_1$ and
\begin{align}
\Psi(N_1,-N_2')= 
 {1\over N_1!}\sum_{s_1,\dots,s_{N_1}= 0}^\infty
 (-1)^{s_1+\cdots+s_{N_1}}
 \prod_{j=1}^{N_1}{(q^{s_j+1})_{N_2'-N_1}\over (-q^{s_j+1})_{N_2'-N_1}}
 \prod_{1\le j<k\le N_1}
 {(1-q^{s_k-s_j})^2\over (1+q^{s_k-s_j})^2}.\label{kadc9Nov12}
\end{align}

The above expression is valid only for $N_2'\ge N_1$.  If $N_2'<N_1$,
then the summation in \eqref{kadc9Nov12} over $N_1$ variables should
reduce to that of $\Zh_{\text{lens}}(N_2',-N_1)_k$ over $N_2'$ variables
to be consistent with the symmetry \eqref{lcqb22Sep12}.  Let us see how
this works by setting $N_2'\to N_2'-\epsilon$ in \eqref{kadc9Nov12}.
Because of \eqref{iamq4Sep12} and \eqref{fxom24Sep12}, only terms that
diverge as $\sim \epsilon^{-(N_1-N_2')}$ in the $s$-sum survive.
Divergences can appear from
\begin{align}
 (q^{s_j+1})_{N_2'-N_1-\epsilon}
 ={1\over (q^{s_j+1+N_2'-N_1-\epsilon})_{N_1-N_2'}}
 ={1\over (1-q^{s_j+1+N_2'-N_1-\epsilon})\cdots (1-q^{s_j-\epsilon})},
\end{align}
where we are keeping only the leading term.
For this to give a divergent $(\sim \epsilon^{-1})$ contribution, it
should be that $s_j+1+N_2'-N_1\le 0$, namely,
$s_j\le N_1-N_2'-1$ (this is impossible for 
$N_1\le N_2'$).  Because $s_1,\dots,s_{N_1}$ should be
different from one another, the most singular case we can have is when
$\{s_1,\dots,s_{N_1}\}\supset \{0,1,\dots,N_1-N_2'-1\}$.  In this case,
we have precisely $\cO(\epsilon^{-(N_1-N_2')})$.  Concretely, let us set
\begin{align}
  s_j&=
 \begin{cases}
  j-1 &\quad (1\le j\le N_1-N_2'),\\
 N_1-N_2'+s'_{j-N_1+N_2'}&\quad
 (N_1-N_2'+1\le j\le N_2')
 \end{cases}\label{uy10Nov12}
\end{align}
with $s_j'\ge 0$ and multiply the result by a combinatoric factor
${N_1\choose N_1-N_2'}\cdot(N_1-N_2')!={N_1!\over N_2'!}$.  By substituting
these into \eqref{kadc9Nov12} and massaging the result, we can show
\begin{align}
 &\Psi(N_1,-N_2'+\epsilon)
 =(-1)^{N_1N_2'+N_1}\left(-{2\over \epsilon \ln q}\right)^{N_1-N_2'}\label{vk10Nov12}\\
 &\qquad
 \times
 {1\over N_2'!}\sum_{s_1',\dots,s_{N_2'}'= 0}^\infty\!\!
 (-1)^{s_1'+\cdots+s_{N_2'}'}
 \prod_{j=1}^{N_2'}{(q^{s_j'+1})_{N_1-N_2'}\over (-q^{s_j'+1})_{N_1-N_2'}}
 \prod_{1\le j<k\le N_2'}{(q^{s_k'-s_j'})_1^2\over (-q^{s_k'-s_j'})_1^2}
 \qquad (N_2'\le N_1).\notag
\end{align}
Namely, the summation over $N_1$ variables $s_1,\dots,s_{N_1}$ correctly
reduced to summation over $N_2'$ variables $s'_1,\dots,s'_{N_2'}$, and
the $\epsilon$ dependence of $\Psi$, combined with $\gamma\sim
\epsilon^{N_1}$, is the correct one to cancel the divergence of $B\sim
\epsilon^{-N_2'}$ (see \eqref{iamq4Sep12}).  So, for $N_2'<N_1$, the
expression for the ABJ partition function 
$\Zh_{\text{ABJ}}(N_1,N_2')_k=\Zh_{\text{lens}}(N_1,-N_2')_k$ is
\begin{align}
 \Zh_{\text{lens}}(N_1,-N_2')_k
 &=
  i^{-{\kappa\over 2}(N_1^2+N_2'^2)}(-1)^{\half N_2'(N_2'-1)}
 2^{-N_2'}(2\pi)^{N_1^2+N_2'^2\over 2}g_s^{N_1+N_2'\over 2}
 q^{-{1\over 6}(N_1-N_2')((N_1-N_2')^2-1)}
\notag\\
&\qquad\qquad
 \times
 (1-q)^{\half(N_1-N_2')(N_1-N_2'-1)}
  B(N_1-N_2',N_1,N_2')
 \Psi(N_2',-N_1)
 \notag\\
 &=
  i^{-{\kappa\over 2}(N_1^2+N_2'^2)}(-1)^{\half N_2'(N_2'-1)}
 2^{-N_2'}(2\pi)^{N_1^2+N_2'^2\over 2}g_s^{N_1+N_2'\over 2}
\notag\\
&\qquad\qquad
 \times
 { q^{-{1\over 6}(N_1-N_2')((N_1-N_2')^2-1)}
 \prod_{j=1}^{N_1-N_2'-1}(q)_j\over G_2(N_1+1)G_2(N_2'+1)}
 \Psi(N_2',-N_1),
 \label{xl10Nov12}
\end{align}
where
\begin{multline}
 \Psi(N_2',-N_1)
 ={1\over N_2'!}\sum_{s_1',\dots,s_{N_2'}'=0}^\infty\!\!
 (-1)^{s_1'+\cdots+s_{N_2'}'}
 \prod_{j=1}^{N_2'}{(q^{s_j'+1})_{N_1-N_2'}\over (-q^{s_j'+1})_{N_1-N_2'}}
 \prod_{1\le j<k\le N_2'}{(q^{s_k'-s_j'})_1^2\over (-q^{s_k'-s_j'})_1^2}
 \qquad (N_2'\le N_1).
 \notag
\end{multline}

Using the explicit expressions \eqref{nijj23Sep12} and
\eqref{xl10Nov12}, It is straightforward to show that the relation
\eqref{lcqb22Sep12} between $\Zh_{\text{lens}}(N_1,-N_2')_k$ and
$\Zh_{\text{lens}}(N_2',-N_1)_{-k}$ holds.

In Table \ref{table:epsilonBehaviorGeneral}, we present a summary of how
various quantities behave as $\epsilon\to 0$ for various values of $N_2$.
\begin{table}
 \begin{center}
 \begin{tabular}{|l||c|cc|cc|c|c|}
  \hline
  \hfil range of $N_2$ \hfil & $B$ & $\beta$ & $\Phi$ & $\gamma$ & $\Psi$ & $S=\beta \Phi=\gamma\Psi$ & $\Zh\propto B S$\\
  \hline\hline
  $N_2>0$ & finite & finite & finite & $\epsilon^{N_1}$ & $\epsilon^{-N_1}$ & finite & finite\\
  \hline
  $N_2<0$, ~$0<N_2'\le N_1$ & $\epsilon^{-N_2'}$ & $\epsilon^{N_2'}$ & finite & $\epsilon^{N_1}$ & $\epsilon^{N_2'-N_1}$ & $\epsilon^{N_2'}$ & finite\\
  \hline
  $N_2<0$, ~$N_1\le N_2'$ & $\epsilon^{-N_1}$ & $\epsilon^{N_1}$ & finite & $\epsilon^{N_1}$ & finite & $\epsilon^{N_1}$ & finite\\
  \hline
 \end{tabular}
 \end{center}
 \caption{\sl The $\epsilon\to 0$ behavior of various quantities for
 general $N_1$.  If $N_2<0$, we define $N_2'=-N_2$.
 \label{table:epsilonBehaviorGeneral}}
\end{table}

%%%%%%%%%%%%%%%%%%%%%%%%%%%%%%%%%%%%%%%%%%%%%%%%%%%%%%%%%%
\section{The perturbative free energy}
\label{Appendix_PE}

In this appendix, we present the free energy of the lens space matrix
model computed by perturbative expansion, up to eight loop order ${\cal
O}(g_s^8)$:
\begin{multline}
 F_{\rm lens}(N_1,N_2)-F^{\rm tree}_{\rm lens}(N_1,N_2)=
 g_s\biggl(\frac{N_1^3}{12}+\frac{N_1^2N_2 }{4} +\frac{ N_1N_2^2}{4}+\frac{N_2^3}{12} -\frac{N_1}{12}-\frac{N_2}{12}\biggr)\\
 +g_s^2\left(\frac{N_1^4}{288}+\frac{N_1^3N_2 }{48} +\frac{N_2^2 N_1^2}{16} +\frac{N_2^3
   N_1}{48}+\frac{N_2^4}{288} -\frac{N_1^2}{288}+\frac{N_1N_2}{48}-\frac{N_2^2}{288}\right)\\
+g_s^4\biggl(-\frac{N_1^6}{86400}-\frac{ N_1^5N_2}{7680}-\frac{N_1^4N_2^2 }{1536}-\frac{5 N_1^3 N_2^3}{1152}-\frac{N_1^2N_2^4 }{1536}-\frac{N_1N_2^5 }{7680}-\frac{N_2^6}{86400} \\
+\frac{N_1^4}{34560}+\frac{7 N_1^3N_2 }{4608}-\frac{ N_1^2N_2^2  }{768}+\frac{7 N_1N_2^3 }{4608}+\frac{N_2^4}{34560}-\frac{N_1^2}{57600}-\frac{N_1N_2}{960}-\frac{N_2^2}{57600}\biggr)\\
 +g_s^6\biggl(\frac{N_1^8}{10160640}+\frac{ N_1^7N_2}{645120}+\frac{N_1^6N_2^2 }{92160}+\frac{N_1^5N_2^3 }{92160}+\frac{7  N_1^4N_2^4}{9216}+\frac{N_1^3N_2^5 }{92160}
+\frac{N_1^2N_2^6 }{92160}+\frac{ N_1N_2^7}{645120}+\frac{N_2^8}{10160640}\\
-\frac{N_1^6}{2177280}+\frac{ N_1^5N_2}{92160}-\frac{N_1^4N_2^2}{2304} +\frac{N_1^3N_2^3 }{27648}-\frac{ N_1^2N_2^4}{2304}+\frac{ N_1N_2^5}{92160}-\frac{N_2^6}{2177280}\\
+\frac{N_1^4}{1451520}+\frac{N_1^3N_2}{11520}+\frac{N_1^2N_2^2}{3840} +\frac{N_1N_2^3}{11520}+\frac{N_2^4}{1451520}-\frac{N_1^2}{3048192}-\frac{N_1N_2}{12096}-\frac{N_2^2}{3048192}\biggr)\\
 +g_s^8\biggl(
 -\frac{N_1^{10}}{870912000}-\frac{17 N_1^9N_2 }{743178240}-\frac{17 N_1^8N_2^2}{82575360}-\frac{N_1^7N_2^3 }{774144}+\frac{97N_1^6N_2^4 }{4423680}-\frac{2821 N_1^5N_2^5 }{14745600} +\frac{97 N_1^4N_2^6}{4423680}\\
-\frac{N_1^3N_2^7 }{774144}-\frac{17 N_1^2N_2^8 }{82575360}-\frac{17 N_1 N_2^9}{743178240}-\frac{N_2^{10}}{870912000}+\frac{N_1^8}{116121600}+\frac{29 N_1^7 N_2}{123863040} -\frac{259 N_1^6 N_2^2}{17694720}\\
+\frac{937 N_1^5 N_2^3}{8847360}+\frac{53 N_1^4 N_2^4}{442368}+\frac{937 N_1^3 N_2^5}{8847360}-\frac{259 N_1^2N_2^6}{17694720}+\frac{29N_1 N_2^7}{123863040}
+\frac{N_2^8}{116121600}-\frac{N_1^6}{41472000} \\
+\frac{853 N_1^5N_2 }{58982400}-\frac{1487 N_1^4N_2^2}{11796480}-\frac{83N_2^3 N_1^3}{1769472} -\frac{1487N_1^2 N_2^4 }{11796480}+\frac{853 N_1 N_2^5}{58982400}-\frac{N_2^6}{41472000}+\frac{N_1^4}{34836480}\\
-\frac{23 N_1^3N_2 }{37158912}+\frac{325N_1^2 N_2^2}{3096576} -\frac{23N_1 N_2^3 }{37158912}+\frac{N_2^4}{34836480}-\frac{N_1^2}{82944000}-\frac{17 N_1N_2}{1382400}-\frac{N_2^2}{82944000}
 \biggr)\ .\nn
\label{feoi21Sep12}
\end{multline}
This perfectly agrees with the result in \cite{Aganagic:2002wv} to the order presented there. 
Meanwhile, we have explicitly checked that the perturbative free energy of the ABJ matrix model is indeed related to the lens space free energy by
\be
 F_{\rm ABJ}(N_1,N_2)= F_{\rm lens}(N_1,-N_2)\ ,
\ee
including the tree contribution with the normalization discussed in Appendix \ref{Appendix_Norm}\@.

%%%%%%%%%%%%%%%%%%%%%%%%%%%%%%%%%%%%%%%%%%%%%%%%%%%%%%%%%%
\section{The Seiberg duality}
\label{Appendix_SD}

In this Appendix, we show that the $(1,N_1)$ ABJ partition function
$Z_{\rm ABJ}(1,N_1)_k$ given in \eqref{ABJN11(2)} is invariant under the
Seiberg duality \eqref{1N2Seibergduality} up to a phase.
Because in the main text we have shown that $Z_{\rm CS}^0(N_2-1)_k$ is
invariant and that the phase factor precisely agrees with the one given
in \cite{Kapustin:2010mh}, all that remains to be shown is the
invariance of the integral $I(1,N_2)_k$ defined in \eqref{1N2integral}.

As claimed in the main text, for Seiberg dual pairs, we can show that
the integrand appearing in $I(1,N_2)_k$ is the same up to a shift in
$s$.  More precisely, the claim to be proven is that the integrand
\begin{align}
 f_{N_2}(s):={\pi\over \sin(\pi s)}\prod_{j=1}^{N_2-1}
 \tan{\pi(s+j)\over |k|}\label{kvze1Dec12}
\end{align}
has the following property:
\begin{align}
 f_{N_2}(s)=f_{\Nt_2}\!\!\left(s-{|k|\over 2}+N_2-1\right),\qquad
 \Nt_2:=2+|k|-N_2.\label{jhyp27Nov12}
\end{align}
Therefore, as long as we take the prescription
\eqref{etaPrescription} for the contour, $I(1,N_2)_k$ defined by the
contour integral \eqref{1N2integral} remains the same.

Note that, if two meromorphic functions $f(s)$ and $g(s)$ have poles and
zeros at the same points and with the same order, then they must be
equal to each other up to an overall constant.  This can be shown as
follows.  If $z=\alpha$ is a pole or a zero, we can write
$f(s)=a(z-\alpha)^n,g(s)=b(z-\alpha)^n$ near $z=\alpha$ by the
assumption.  This means that $f'/f=g'/g =n(z-\alpha)^{-1}$ near
$z=\alpha$.  Now, recall that Mittag-Leffler's theorem in complex
analysis states that, if two functions have poles at the same points and
if the singular part of the Laurent expansion around each of them is the
same, then the two functions are identical.  So, because $f'/ f$ and
$g'/ g$ share poles and residues, they must be identical. This means
that $f(s)=cg(s)$ with a constant $c$.  In the present case, it is easy
to show that the overall scale of $f_{N_2}(s)$ and $f_{\Nt_2}(s-{|k|\over
2}+N_2-1)$ is the same asymptotically, because both tend to ${2\pi
i^{N_2-2}e^{-\pi \sigma}}$ for $s=i\sigma$, $\sigma\to +\infty$.  So, in
order to show that these two functions are equal, we only have to show
that they share poles and zeros.

So, let us compare the poles and zeros of the two functions $f_{N_2}(s)$
and $f_{\Nt_2}(s-{|k|\over 2}+N_2-1)$.  Recall the expression for
$f_{N_2}(s)$ given by \eqref{kvze1Dec12}.  First, ${\pi\over \sin(\pi
s)}$ gives simple poles at $s\in \bbZ$ (P poles) but no zero.  On the
other hand, $\tan{\pi (s+j)\over |k|}$ gives simple poles at
$s=|k|(p+\half)-j$, $p\in\bbZ$ (NP poles), and simple zeros at $s=|k|q-j$,
$q\in\bbZ$ (NP zeros).  Using this data, we can find the pole/zero
structure of the two functions as we discuss now.  We should consider
odd and even $k$ cases separately,

\begin{figure}[h!]
\begin{center}
\begin{tabular}{cc}
\includegraphics[height=5.6cm]{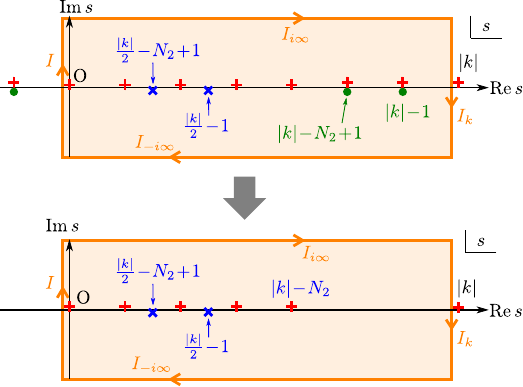}
&
\includegraphics[height=5.6cm]{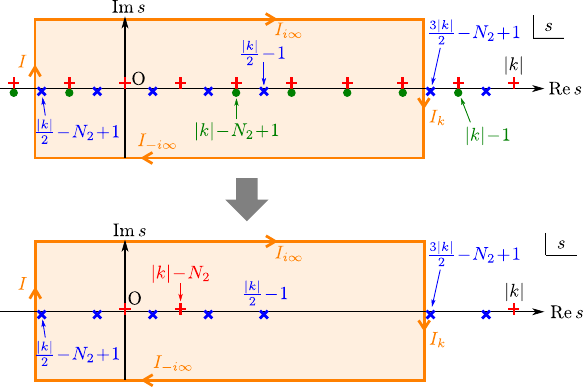}
\\
 \begin{minipage}{7cm}
  \begin{itemize}
   \item[(a)] ${|k|\over 2}-N_2+1\ge 1$ (shown is the $N_2=3,|k|=7$ case).
  \end{itemize}
   \end{minipage}
 &
 \begin{minipage}{7cm}
  \begin{itemize}
   \item[(b)] ${|k|\over 2}-N_2+1\le 1$ (shown is the $N_2=6,|k|=7$ case).
  \end{itemize}
 \end{minipage}
\end{tabular}
\caption{\sl The pole/zero structure of the integrand function for odd $k$.
``$+$'' (red) denotes the P pole, ``$\times$'' (blue) the NP pole, and
``$\bullet$'' (green) the NP zero.  Some poles and zeros are shown
slightly above or below the real $s$ axis, but this is for the
convenience of presentation and all poles and zeros are on the real $s$
axis.  If the original theory is in case (a) ${|k|\over 2}-N_2+1\ge 1$,
then the Seiberg dual is in case (b) ${|k|\over 2}-N_2+1\le 1$, and {\it
vice versa.} In the figure, actual Seiberg dual theories are shown.  In
the upper panels, all P poles coming from ${\pi\over \sin \pi s}$ and
all NP poles and NP zeros coming from $\prod_j \tan$ are shown.  In the
lower panels, P poles and NP zeros that cancel each other are removed.
We see that the actual poles are the same in the dual theories (a) and
(b),  with P and NP poles interchanged.
\label{poles_gen_odd}}
\end{center}
\end{figure}

\paragraph{Odd $\boldsymbol{k}$:} For odd $k$, $f_{N_2}(s)$ has poles but no zeros.
All poles are simple
poles and they can be divided into two groups:
\begin{align}
\begin{split}
 	\rm P &:\qquad s=0,\dots,|k|-N_2,\\
	\rm NP&:\qquad s={|k|\over 2}-N_2+1,\dots,{|k|\over 2}-1,
\end{split}\label{itpg27Nov12}
\end{align}
where periodicity $s\cong s+|k|$ is understood;
see Figure \ref{poles_gen_odd}.
Note that this is valid
even for ${|k|\over 2}-N_2+1<0$, for which some of the poles are at $s<0$.  P means poles coming from $\pi \over \sin\pi s$ while NP
means poles coming from $\prod_j\tan$.  Some of the P poles got canceled
by NP zeros and reduced to regular points.  NP poles are not canceled.
P and NP poles never collide, because the former are at integral $s$
while the latter are at half-odd-integral $s$.

\eqref{itpg27Nov12} means that $f_{\Nt_2}(s)$ has simple poles at
\begin{align}
\begin{split}
 	\rm  P&:\qquad s=0,\dots,|k|-\Nt_2=0,\dots,-2+N_2,\\
	\rm NP&:\qquad s={|k|\over 2}-N_2+1,\dots,{|k|\over 2}-1
 =-{|k|\over 2}+N_2-1,\dots,{|k|\over 2}-1,
\end{split}
\end{align}
which in turn means that $f_{\Nt_2}(s+N_2-{|k|\over 2}-1)$ has simple
poles at
\begin{align}
\begin{split}
 	\rm  P&:\qquad s={|k|\over 2}-N_2+1,\dots,{|k|\over 2}-1,\\
	\rm NP&:\qquad s=0,\dots,|k|-N_2;
\end{split}
\end{align}
This is the same as \eqref{itpg27Nov12}, with P and NP interchanged.
This proves the identity \eqref{jhyp27Nov12} for odd $k$.  Figure
\ref{poles_gen_odd} shows the explicit pole/zero structure in the specific
case of $U(1)_7\times U(3)_{-7}=U(1)_{-7}\times U(6)_{7}$.

\begin{figure}[h!]
\begin{center}
\begin{tabular}{cc}
\includegraphics[height=5.1cm]{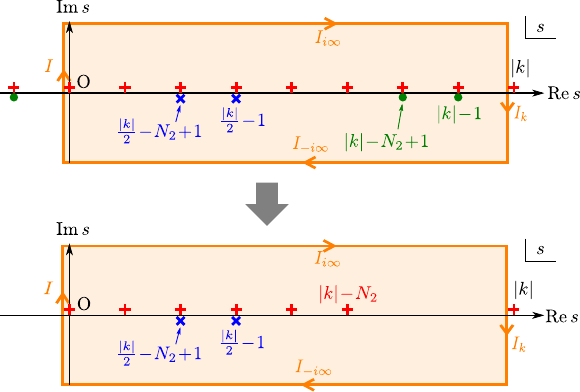}
&
\includegraphics[height=5.1cm]{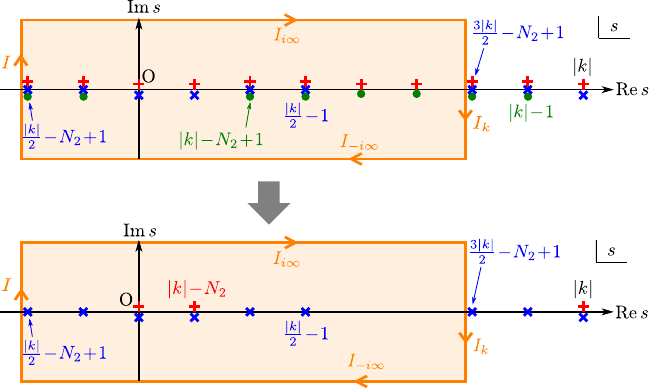}
\\
 \begin{minipage}{7cm}
  \begin{itemize}
   \item[(a)] ${|k|\over 2}-N_2+1\ge 1$ (shown is the $N_2=3,|k|=8$ case).
  \end{itemize}
   \end{minipage}
 &
 \begin{minipage}{7cm}
  \begin{itemize}
   \item[(b)] ${|k|\over 2}-N_2+1\le 1$ (shown is the $N_2=7,|k|=8$ case).
  \end{itemize}
 \end{minipage}
\end{tabular}
\caption{\sl The pole/zero structure of the integrand function for even
$k$.  See Figure \ref{poles_gen_odd} for explanation of the symbols.  In
the figure, actual Seiberg dual theories are shown.  In the upper
panels, all P poles coming from ${\pi\over \sin \pi s}$ and all NP poles
and NP zeros coming from $\prod_j \tan$ are shown.  In the lower panels,
poles that are canceled by NP zeros are removed.  If a P pole, a NP pole
and a NP zero all collide, the resulting simple pole is interpreted as a
NP pole.  The surviving poles are the same in the dual theories (a) and
(b), with P and NP poles interchanged.  \label{poles_gen_even} }
\end{center}
\end{figure}

\paragraph{Even $\boldsymbol{k}$:} Also for even $k$, the function $f_{N_2}(s)$ has poles but no zeros.
Some of the poles are simple while others are double.  Let us think of a
double pole as made of two simple poles on top of each other.  Then
there are two groups of simple poles, as follows:
\begin{align}
\begin{split}
 	\rm  P&:\qquad s=0,\dots,|k|-N_2,\\
	\rm NP&:\qquad s={|k|\over 2}-N_2+1,\dots,{|k|\over 2}-1,
\end{split}\label{kabs27Nov12}
\end{align}
where $s\cong s+|k|$ is again implied;
see Figure \ref{poles_gen_even}.
For $k$ even, NP zeros can cancel
P poles and NP poles, and it becomes ambiguous whether we should call a
particular pole P or NP\@.  This happens in the ${|k|\over 2}-N_2+1<0$
case, where a P pole, a NP pole and a NP zero all can be at the same
point.  When this happens, we think of the P pole getting canceled by
the NP zero, and group the remaining simple pole into NP, as we did above.  This
is arbitrary, but it is a unique choice for which the structure
\eqref{kabs27Nov12} becomes identical to the odd $k$ case,
\eqref{itpg27Nov12}.

Because \eqref{kabs27Nov12} is the same as the odd $k$ case,
\eqref{itpg27Nov12}, the rest goes exactly the same, and we conclude
that $f_{N_2}(s)$ and $f_{\Nt_2}(s+N_2-{|k|\over 2}-1)$ are identical,
with P and NP interchanged.
Figure \ref{poles_gen_even} shows the explicit pole/zero structure in
the specific case of $U(1)_8\times U(3)_{-8}=U(1)_{-8}\times U(7)_{8}$.

\end{document}